\documentclass[fleqn,usenatbib]{mnras}

\usepackage{newtxtext,newtxmath}

\usepackage[T1]{fontenc}

\DeclareRobustCommand{\VAN}[3]{#2}
\let\VANthebibliography\thebibliography
\def\thebibliography{\DeclareRobustCommand{\VAN}[3]{##3}\VANthebibliography}

\usepackage{graphicx}
\usepackage{amsmath}	
\usepackage{caption}
\usepackage{subcaption}
\usepackage{threeparttable}
\usepackage{arydshln}
\usepackage{float}
\usepackage{multirow}

\newcommand{\msun}{M$_\odot$}
\newcommand{\rsun}{R$_\odot$}
\newcommand{\kms}{$\mathrm{km}\,\mathrm{s}^{-1}$}
\newcommand{\ms}{$\mathrm{m}\,\mathrm{s}^{-1}$}
\newcommand{\unsim}{\mathord{\sim}}
\newcommand{\chisqr}{$\chi^2_r$}

\title[The T Tauri star V410 Tau in the eyes of SPIRou and TESS]{The T Tauri star V410 Tau in the eyes of SPIRou and TESS}

\author[B. Finociety et al.]{B. Finociety$^{1}$ \thanks{E-mail: benjamin.finociety@irap.omp.eu},
J.-F. Donati$^{1}$, B. Klein$^{1,2}$, B. Zaire$^{1}$, L. Lehmann$^{1}$, C. Moutou$^{1}$, J. Bouvier$^{3}$, S.H.P Alencar$^{4}$,\newauthor L. Yu$^{5}$, K. Grankin$^6$, É. Artigau$^{7}$, R. Doyon$^{7}$, X. Delfosse$^{3}$, P. Fouqué$^1$, G. Hébrard$^8$, M. Jardine$^9$, Á. Kóspál$^{10}$,\newauthor F. Ménard$^3$ and the SLS consortium
\\
$^{1}$ Université de Toulouse, CNRS, IRAP, 14 av. Belin, 31400 Toulouse, France \\
$^{2}$ Sub-department of Astrophysics, Department of Physics, University of Oxford, Oxford OX1 3RH, UK \\
$^{3}$ Université Grenoble Alpes, CNRS, IPAG, 38000 Grenoble, France \\
$^{4}$ Departamento de Fisica – ICEx – UFMG, Av. Antonio Carlos 6627, 30270-901 Belo Horizonte, MG, Brazil \\
$^{5}$ CNES, 18 avenue Edouard Belin, 31401, Toulouse, France \\
$^6$ Crimean Astrophysical Observatory, 298409, Nauchny, Crimea \\
$^7$ Observatoire du Mont-Mégantic, Département de physique de l'Université de Montréal, iREx, Montréal, Canada \\
$^8$ Institut d’astrophysique de Paris, UMR7095 CNRS, Université Pierre \& Marie Curie, 98bis boulevard Arago, 75014 Paris, France \\
$^9$ SUPA, School of Physics and Astronomy, University of St Andrews, North Haugh, KY16 9SS, UK \\
$^{10}$ Konkoly Observatory, Research Centre for Astronomy and Earth Sciences, Konkoly-Thege Mikl\'os \'ut 15-17, 1121 Budapest, Hungary \\
}

\date{Accepted 2021 September 23. Received 2021 September 23; in original form 2021 June 4.}

\pubyear{2021}

\begin{document}
\label{firstpage}
\pagerange{\pageref{firstpage}--\pageref{lastpage}}
\maketitle

\begin{abstract}
We report results of a spectropolarimetric and photometric monitoring of the weak-line T Tauri star V410~Tau based on data collected mostly with SPIRou, the near-infrared (NIR) spectropolarimeter recently installed at the Canada-France-Hawaii Telescope, as part of the SPIRou Legacy Survey large programme, and with TESS between October and December 2019.
Using Zeeman-Doppler Imaging (ZDI), we obtained the first maps of photospheric brightness and large-scale magnetic field at the surface of this young star derived from NIR spectropolarimetric data. For the first time, ZDI is also simultaneously applied to high-resolution spectropolarimetric data and very-high-precision photometry. V410~Tau hosts both dark and bright surface features and magnetic regions similar to those previously imaged with ZDI from optical data, except for the absence of a prominent dark polar spot. The brightness distribution is significantly less contrasted than its optical equivalent, as expected from the difference in wavelength.  
The large-scale magnetic field ($\unsim410$~G), found to be mainly poloidal, features a dipole of $\unsim390$~G, again compatible with previous studies at optical wavelengths.
NIR data yield a surface differential rotation slightly weaker than that estimated in the optical at previous epochs.
Finally, we measured the radial velocity of the star and filtered out the stellar activity jitter using both ZDI and Gaussian Process Regression down to a precision of  $\unsim0.15$ and 0.08~\kms\ RMS, respectively, confirming the previously published upper limit on the mass of a potential close-in massive planet around V410~Tau.

\end{abstract}

\begin{keywords}
techniques: polarimetric -- stars: imaging -- stars: individual: V410 Tau -- stars: magnetic field -- stars: activity
\end{keywords}



\section{Introduction}

Young low-mass stars ($<2$~\msun), aged of 1-15 Myr, that have just emerged from their dust cocoon and are still contracting towards the main sequence are called T~Tauri stars (TTSs). These pre-main sequence (PMS) stars are divided into two classes: the classical T~Tauri stars (cTTSs), still surrounded by an accretion disc and accreting from its inner regions, and the weak-line T~Tauri stars (wTTSs) that are no longer accreting and have exhausted the inner regions of their discs (or the whole disc). These stars are of obvious interest for further constraining theoretical models of star/planet formation, especially considering the role that stellar magnetic fields play at early stages of evolution \citep{bouvier07a}.

Magnetospheric accretion/ejection processes at work in cTTSs have been studied with increasing attention since the first detection of magnetic fields in such stars by \cite{johnskrull99}: e.g. AA Tau \citep{bouvier07b}, V2129 Oph \citep{donati07,donati11,alencar12}, DN Tau \citep{donati13}, LkCa 15 \citep{alencar18,donati19}, HQ Tau \citep{pouilly20} or DoAr44 \citep{Bouvier20}. Most of these studies were made possible thanks to the MaPP (Magnetic Protostars and Planets) Large Observing Programme, dedicated to the observation of magnetized PMS stars and their accretion discs, carried out with the ESPaDOnS high-resolution spectropolarimeter on the 3.6~m Canada-France-Hawaii Telescope (CFHT). These studies suggested that the magnetic topologies of cTTSs mainly reflect the internal stellar structure as they do for main-sequence (MS) low-mass stars \citep{donati09,morin10,gregory12}.

The MaTYSSE (Magnetic Topologies of Young Stars and the Survival of close-in giant Exoplanets) Large Programme, also carried out mostly with ESPaDOnS at the CFHT, was dedicated to the observation of wTTSs with the aim of investigating how different magnetic fields of wTTSs are from those of cTTSs, and whether/how frequently they host close-in massive planets (hot Jupiters/hJs). 
A few tens of wTTSs have been studied within MaTYSSE, for example LkCa 4 \citep{donati14}, V830 Tau \citep{donati17}, TAP 26 \citep{yu17}, Par 1379 and Par 2244 \citep{hill17}, V410 Tau \citep{yu19}. In all cases, the large-scale magnetic field of the star was mapped using Zeeman-Doppler Imaging (ZDI), a tomographic technique inspired from medical imaging which proved very efficient at recovering stellar magnetic topologies \citep{semel89,brown91,donatibrown97,donati06}. This technique revealed the wide diversity of magnetic field topologies that can be found on wTTSs. Whereas most wTTSs show the same magnetic trends as those reported for cTTSs, some of them depart from this picture, with, e.g., V410~Tau showing a strong azimuthal field despite being fully convective. MaTYSSE also enabled the detection of hJs around 2 wTTSs \citep{donati17,yu17} through the periodic velocimetric signal they induce in the spectral lines of their host stars.

The SPIRou Legacy Survey (SLS) is a new Large Programme allocated on CFHT with SPIRou, the new cryogenic spectropolarimeter/high-precision velocimeter operating at near-infrared wavelengths (0.95-2.55 $\mu$m, \citealt{donati20b}).  The SLS includes in particular a work package focusing on cTTSs and wTTSs, with the goal of studying further their magnetic topologies and the potential presence of hJs.  Infrared wavelengths are indeed well adapted for measurements of stellar magnetic fields thanks to the enhanced Zeeman effect.  The impact of activity (and in particular of surface brightness features) on the shape of line profiles, and thereby on the measured radial velocity, is also expected to be smaller \citep{mahmud11,crockett12}, making it easier to ascertain the presence of potential hJs.  

In this paper, we report the results about V410 Tau, obtained in the framework of SLS. V410 Tau is a triple star system with the main star V410 Tau A being much brighter than the two other companions \citep{ghez97}. In particular, \cite{ghez97} reported a difference of $\unsim$2.5~mag (resp. $\unsim3$~mag) between V410~Tau~A and B (resp. C) in the $K$ band. From the brightness measurements collected by \cite{ghez97} and the colour indexes for young PMS stars derived by \cite{pecaut13}, we also estimated a difference of 3~mag (resp. 5~mag) between V410~Tau~A and B (resp. C) in the $I_{\rm c}$ band. 

V410~Tau is a young, fully convective wTTS with an age of about 1~Myr hosting a complex magnetic field \citep{skelly10,yu19}. Located at a distance of $129.4\pm0.4$~pc in the Taurus star forming complex, V410~Tau belongs to the youngest substellar region C2-L1495 whose age was recently estimated from GAIA data ($1.34\pm0.19$~Myr, \citealt{krolikowski21}). V410~Tau has an effective temperature $T_{\rm eff}$ and a logarithmic gravity of 4500~K and 3.8 (in cgs units), respectively, for a mass of $1.42\pm0.15$~\msun\ and a radius of $3.40\pm0.5$~\rsun\ \citep{yu19}. Recent stellar models (e.g. \citealt{baraffe15}) suggest an even younger age (<0.5~Myr, hardly compatible with results from GAIA) and a lower mass ($1.14\pm0.10$~\msun) as mentioned in \cite{yu19}. All models suggest that V410~Tau is a fully-convective star.

\cite{yu19} performed a thorough spectropolarimetric analysis of V410 Tau based on optical data collected between 2008 and 2016. This study revealed in particular that the brightness distribution and the large-scale magnetic topology at the surface of the star evolve with a timescale of months (160 to 600~d) but also that the surface is sheared by a weak level of differential rotation ($9.7\pm0.3~\rm mrad\ d^{-1}$). The magnetic field of V410~Tau exhibits a strong toroidal component whose contribution to the overall magnetic energy decreased from $\unsim50\%$ in 2008 to $\unsim30\%$ in 2016. In addition, the strength of the dipole followed the opposite trend, increasing from $\unsim130$~G to $\unsim400$~G. Although their data were spread over several years, they were not able to identify a clear magnetic cycle but only a lower limit (of $\simeq$8~yr) for its duration (if a cycle is indeed present). The study of radial velocities coupled to stellar surface imaging revealed that no planet more massive than 1~M$_{\rm jup}$ orbits the host star within a distance of 0.09~au.

V410~Tau has been monitored both with SPIRou from 2019 October 31 to December 13, and with the Transiting Exoplanet Survey Satellite (TESS) from 2019 November 28 to December 23 during its monitoring of sector 19. Additional contemporaneous ground-based photometric data were also collected at the Crimean Astrophysical Observatory (CrAO) over the same observing season.  
We start this paper with the description (in Section \ref{sec:sec2}) of both spectropolarimetric and photometric observations collected for our study. In Section \ref{sec:sec3}, we report our investigations about the surface brightness and the large-scale magnetic field with ZDI. In Section \ref{sec:sec4} we investigate the stellar activity both with radial velocity (RV) measurements and activity indicators (based on the \ion{He}{i} triplet at 1083.3~nm and the Paschen $\beta$ and Brackett $\gamma$ lines). Finally, we summarize and discuss our main results in Section \ref{sec:sec5}.

\section{Observations}
\label{sec:sec2}
\subsection{SPIRou observations}
\label{sec:sec2.1}
We performed spectropolarimetric observations with SPIRou between 2019 October 31 and 2019 December 13 (UTC). SPIRou works in the near-infrared (NIR) domain between 950~nm and 2500~nm with a spectral resolving power of $\unsim$70 000 \citep{donati20b}. Each polarimetric observation is composed of a sequence of 4 subexposures of 300 seconds each taken with different configurations of the polarimeter (i.e., different azimuths of the polarimeter retarders) in order to get rid of potential spurious signals in the polarisation and systematic errors at first order \citep{donati97}. Twenty sequences were collected, yielding spectra in both unpolarized (Stokes~$I$) and circularly polarized (Stokes $V$) light. Data reduction was performed with a pipeline based on the ESPaDOnS pipeline Libre-ESpRIT \citep{donati97} adapted for SPIRou observations. We then obtained telluric-corrected spectra using a PCA approach as mentionned by \cite{artigau14}.
The signal-to-noise ratio (SNR) per pixel of these spectra peaks in the $H$ band with a median value of 179 (ranging from 140 to 200). We applied Least-Square Deconvolution (LSD; \citealt{donati97}) on all spectra in order to add up information from all lines. 
We only used 18 out of the 20 recorded spectra, the two remaining ones (collected on 2019 Nov 10 and Dec 13) being either much noiser than the average or suffering from a technical issue. 

Three series of LSD profiles were computed with three different masks generated with the VALD-3 database \citep{vald} and including lines ranging from 950 to 2600~nm. The first one (hereafter M1) contains $\unsim$10~000 atomic and molecular lines. It yields Stokes~$I$ LSD profiles only, as the magnetic sensitivity (i.e. Landé factor) of many molecular lines is unknown, with SNRs ranging from 1890 to 3060 (median of 2770). The second mask (hereafter M2) contains $\unsim2000$ atomic lines with known Landé factors and relative depths with respect to the continuum >10\%. It yields Stokes~$I$ LSD profiles with SNRs ranging from 1270 ro 1930 (median of 1740) and Stokes~$V$ LSD profiles with SNRs ranging from 3380 to 4950 (median of 4440). The last mask (M3 hereafter), containing $\unsim900$ molecular lines with a depth relative to the continuum down to 5\%, yields Stokes~$I$ LSD profiles with SNRs ranging from 600 to 1020 (median of 835).

To phase our observations on the rotation cycle of the star, we used the same reference date as in \cite{skelly10} and \cite{yu19}, namely the Barycentric Julian Date BJD$_0$=2,454,832.58033, along with the well defined stellar rotation period obtained by \cite{stelzer03}: $P_{\mathrm{rot}}~=~1.871970~\pm~0.000010$~d. Despite the very well constrained rotation period, the reconstructed maps presented in Sec.~\ref{sec:sec3} cannot be directly compared to the previously published ones, as the brightness and magnetic surface maps of V410~Tau evolved since then as a result of both differential rotation and intrinsic variability. 

A summary of our observations is given in Table \ref{table1}. Since our data are spread over 23 rotation cycles only (i.e. 43~d), we do not expect major changes in the surface brightness map nor in the large-scale magnetic topology of V410~Tau, given the typical timescale on which both quantities are found to evolve, either as a result of intrinsic variability or differential rotation \citep{yu19}.

\addtolength{\tabcolsep}{6pt}  
\begin{table*}
\caption{Spectropolarimetric observations of V410~Tau obtained with SPIRou between 2019 October and December. Each observation is composed of a sequence of four 300~s-subexposures. The 1$^{\rm st}$ to the 4$^{\rm th}$ columns give the date, the Coordinated Universal Time, the Barycentric Julian Date and the rotation cycle (computed as indicated in Sec.~\ref{sec:sec2.1}). Columns 5 to 9 list the SNRs of the spectra in the $H$ band, in the Stokes~$I$ LSD profiles provided by masks M1, M2 and M3 and in the Stokes~$V$ LSD profiles obtained with mask M2. From column 10 to 14, we detail the measured RV, the longitudinal magnetic field, and the activity indicators (see Sec.~\ref{sec:activity_index}) based on the \ion{He}{i} triplet at 1083.3~nm, Paschen $\beta$ and Brackett $\gamma$ lines, along with their error bars estimated from photon noise only (or taking into account intrinsic variability as well for the number in parenthesis, see Sec.~\ref{sec:activity_index}).}
\label{table1}
\centering 
\resizebox{\textwidth}{!}{
\begin{threeparttable}
\begin{tabular}{lccccccccccccc}
\\
\hline \hline
\multicolumn{1}{c}{Date} & UTC & BJD & Cycle & SNR & \multicolumn{3}{c}{SNR$_I$} & SNR$_V$ & RV  & B$_l$ & \multicolumn{3}{c}{Activity proxies}\\[1mm]
\multicolumn{1}{c}{} &  &  &  & & (M1) & (M2) & (M3) & (M2) &  &  & \ion{He}{i} & Pa$\beta$ & Br$\gamma$ \\ 
\multicolumn{1}{c}{2019} &  & 2458700+ & 2112+ & &  &  & &  & (\kms) & (G) & (nm) & (nm) & (nm) \\ \hline

Oct 31 & 12:07:55 & 88.012 & 0.978 & 179 & 1890 & 1270 & 600 & 4950 & -1.923 $\pm$ 0.273 & $-182\pm29$  & $0.129\pm0.002$ (0.025) & $0.009\pm0.001$ (0.003) & $-0.003\pm0.004$ (0.006) \\
Nov 01 & 12:16:25 & 89.018 & 1.516 & 193 & 2070 & 1380 & 600 & 4460 & 0.901 $\pm$ 0.258 & $-195\pm32$ & $-0.152\pm0.002$ (0.025) & $-0.005\pm0.001$ (0.003) & $-0.009\pm0.004$ (0.006) \\ 
Nov 03 & 10:43:17 & 90.954 & 2.550 & 202 & 2890 & 1820 & 890 & 4870 & 1.149 $\pm$  0.177 & $-224\pm29$ & $-0.202\pm0.002$ (0.025) & $-0.007\pm0.001$ (0.003) & $-0.008\pm0.004$ (0.006)\\
Nov 04 & 14:01:42 & 92.092 & 3.157 & 185 & 2570 & 1680 & 710 &4440 & -1.589 $\pm$  0.202 & $-83\pm32$ & $0.113\pm0.002$ (0.025) & $-0.002\pm0.001$ (0.003) & $-0.000\pm0.004$ (0.006)\\
Nov 05 & 11:23:26 & 92.981 & 3.633 & 197 & 2880 & 1830 & 870 & 4700 & 2.292 $\pm$  0.175 & $-198\pm30$ & $-0.008\pm0.002$ (0.025) & $-0.001\pm0.001$ (0.003) & $-0.005\pm0.004$ (0.006)\\
Nov 07 & 10:53:20 & 94.961 & 4.690 & 171 & 2760 & 1730 & 760 & 3870 & 2.553 $\pm$  0.179 & $-252\pm37$ & $0.070\pm0.002$ (0.025) & $0.005\pm0.001$ (0.003) & $0.007\pm0.004$ (0.006) \\
Nov 08 & 12:48:55 & 96.041 & 5.267 & 187 & 3050 & 1930 & 860& 4400 & -0.302 $\pm$  0.175 & $-15\pm32$ & $0.096\pm0.002$ (0.025) & $-0.006\pm0.001$ (0.003) & $-0.004\pm0.004$ (0.006) \\
Nov 09 & 10:01:53 & 96.925 & 5.739 & 176 & 2070 & 1400 & 660& 4560 & 2.349 $\pm$  0.245 & $-148\pm31$ & $-0.014\pm0.002$ (0.025) & $0.006\pm0.001$ (0.003) & $0.009\pm0.004$ (0.006)\\
Nov 10$^{(a)}$ & 11:34:51 & 97.988 & 6.305 & 143 & - & - & -& - & -  & - & - & - & -\\
Nov 11 & 11:19:14 & 98.979 & 6.837 & 197 & 2660 & 1680 & 810 &4370 & 0.294 $\pm$  0.193 & $-91\pm33$  & $-0.011\pm0.002$ (0.025) & $0.005\pm0.001$ (0.003) & $-0.003\pm0.004$ (0.006)\\
Nov 13 & 09:37:13 & 100.908 & 7.867 & 179 & 2720 & 1720 & 820 & 3950 & -0.128 $\pm$  0.192 & $-142\pm36$ & $0.015\pm0.002$ (0.025) & $0.005\pm0.001$ (0.003) & $0.000\pm0.004$ (0.006)\\
Nov 14 & 10:00:42 & 101.924 & 8.410 & 172 & 2710 & 1720 & 850 & 4210 & 0.627 $\pm$  0.197 & $-115\pm34$ & $-0.075\pm0.002$ (0.025) & $-0.001\pm0.001$ (0.003) & $0.004\pm0.004$ (0.006) \\

Dec 05 & 09:34:52 & 122.906  & 19.619 & 155 & 2780 & 1760 & 760 & 3580 & 2.127 $\pm$  0.182 & $-223\pm40$ & $-0.057\pm0.002$ (0.025) & $0.002\pm0.001$ (0.003) & $-0.004\pm0.004$ (0.006)   \\
Dec 07 & 10:50:36 & 124.960 & 20.715 & 150 & 2320 & 1550 & 690 & 3380 & 2.767 $\pm$  0.213 & $-211\pm42$ & $0.028\pm0.002$ (0.025) & $0.002\pm0.001$ (0.003) & $0.005\pm0.004$ (0.006) \\
Dec 08 & 09:11:20 & 125.890 & 21.213 & 175 & 2940 & 1870 & 930 & 4120 & -1.043 $\pm$  0.179 & $-33\pm35$ & $0.099\pm0.002$ (0.025) & $0.003\pm0.001$ (0.003) & $0.002\pm0.004$ (0.006)\\
Dec 09$^{(b)}$ & 09:25:30 & 126.900 & 21.752 & 191 & 3000 & 1900 & 1020 &4570 & 1.851 $\pm$  0.171 & $-131\pm31$& - & - & -\\
Dec 10 & 08:16:38 & 127.852 & 22.261 & 186 & 3060 & 1910 & 1020 &4750 & -0.423 $\pm$  0.165 & $47\pm30$ & $0.114\pm0.002$  & $0.014\pm0.001$  & $0.012\pm0.004$\\
Dec 11 & 11:06:22 & 128.970 & 22.858 & 197 & 2870 & 1800 & 970 & 4620 & 0.060 $\pm$  0.180 & $-87\pm31$ & $0.172\pm0.002$  & $0.026\pm0.001$  & $0.009\pm0.004$\\
Dec 12 & 09:25:49 & 129.900 & 23.355 & 176 & 2920 & 1830 & 910 & 4440 & 0.377 $\pm$  0.177 & $23\pm32$ & $-0.045\pm0.002$ (0.025) & $-0.006\pm0.001$ (0.003) & $0.012\pm0.004$ (0.006)\\ 
Dec 13$^{(a)}$ & 09:32:01 & 130.904 & 23.891 & 166 & - & - & - & - & -  & - & - & - & - \\ \hline

\end{tabular}

\begin{tablenotes}
\item $^{(a)}$ Removed from the analysis (see Sec.~\ref{sec:sec2.1})
\item $^{(b)}$ Removed from the activity indicators due to a flare (see Sec.~\ref{sec:activity_index})
\end{tablenotes}

\end{threeparttable}
}
\end{table*}
\addtolength{\tabcolsep}{-6pt} 

\subsection{TESS observations}
\label{sec:sec2.2}
Contemporaneously with SPIRou observations, V410~Tau (TIC 58231482) was observed by TESS \citep{ricker14} from 2019 November 28 to 2019 December 23 during its monitoring of sector 19. TESS being mostly sensitive to wavelengths in the range 600 to 1000~nm (centred on the Cousins $I_{\rm c}$ photometric band), it thus probes a different surface brightness distribution than that observed with SPIRou. 

\begin{figure}
    \centering 
    \includegraphics[scale=0.19]{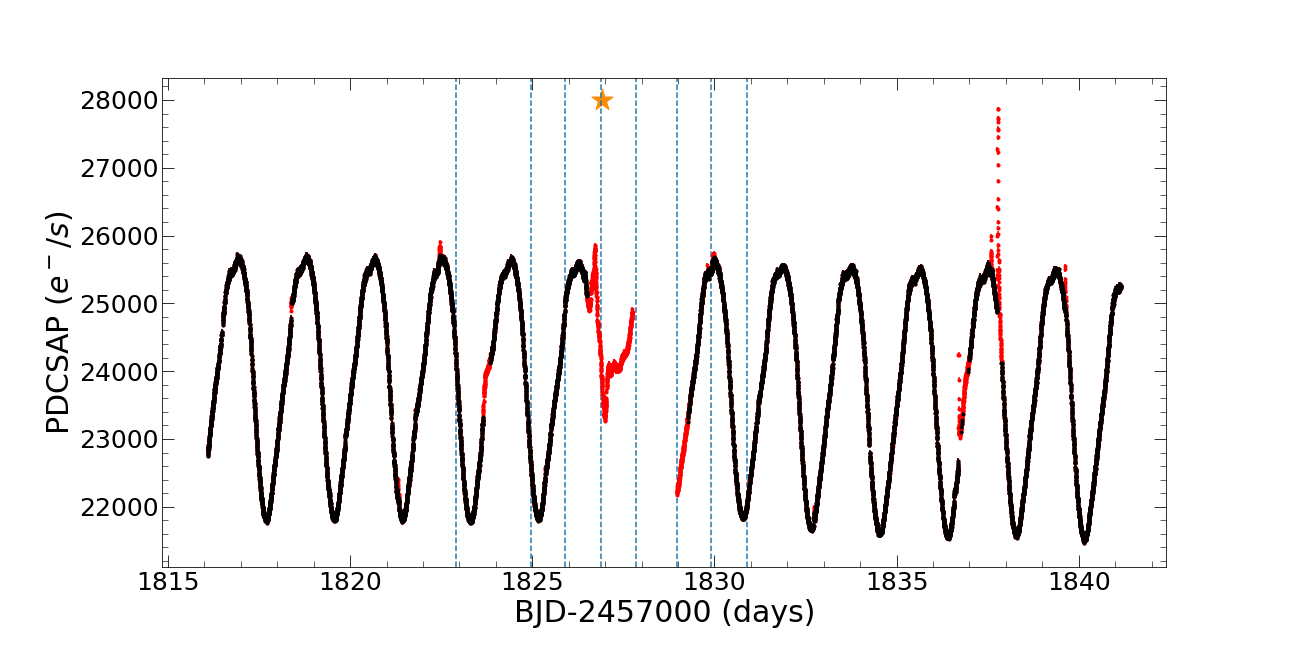}
    \caption{TESS Pre-search Data Conditioning Single Aperture Photometry. The black dots are the ones used in the study, while the red ones have been flagged by the SPOC pipeline or rejected because of the high background level or filtered by a 3$\sigma$-clipping process (see Sec. \ref{sec:sec2.2}) and have thus not been taken into account in this work. 
    The dashed vertical lines depict SPIRou observations contemporary with the TESS monitoring (additional SPIRou observations were collected before). The orange star flags the observation affected by a flare (see Sec.~\ref{sec:activity_index}).}
    \label{fig:Lc_tess}
\end{figure}

The target was observed with a cadence of 2 min over a total time span of 25~d. These observations have been re-processed by the Science Processing Operations Center (SPOC; \citealt{jenkins16}) data pipeline (version 4.0), with light curves available from the Mikulski Archive for Space Telescopes (MAST)\footnote{\url{https://archive.stsci.edu/missions-and-data/tess}}. In particular, we used the Pre-search Data Conditioning Single Aperture Photometry (PDCSAP) flux that provides a better estimate of the intrinsic variability of the star since instrumental variations have been corrected in this light curve, as well as contamination from some nearby stars \citep{smith12,stumpe12,stumpe14}. We only kept the data that were not flagged from the SPOC pipeline, thus those with no known problems. We also rejected the observations carried out between BJD 2,458,826.5 and 2,458,827.8, for which a high background level from the Earth led to spurious photometric variations\footnote{Details can be found in the Data Release notes of sector 19 (DR26 and DR30) available at \url{https://archive.stsci.edu/tess/tess_drn.html}} (see Fig. \ref{fig:Lc_tess}).

Since TESS is sensitive to small flux variations caused by flares, we first filtered them through a $3\sigma$-clipping process, which also allowed us to mitigate the potential effects of activity from stellar companions (V410~Tau~B and C) on the photometry. This process consisted in an iterative fit of the light curve with a Gaussian Process Regression (GPR; \citealt{rasmussen06}). At each iteration, we rejected the points having a residual larger than 3$\sigma$ until all the residuals were lower than this threshold.
As we expected that the variations in the light curve are mainly due to stellar activity, we chose a quasi-periodic kernel to model the TESS light curve, which is well adapted to model signals induced by active regions rotating with the stellar disk \citep{haywood14}:%
\begin{equation}
    k(t_i,t_j) = \theta_1^2 \exp\left[-\frac{(t_i-t_j)^2}{\theta_2^2} - \frac{\sin^2\frac{\pi(t_i-t_j)}{\theta_3}}{\theta_4^2}\right],
    \label{eq1}
\end{equation}
where $t_i, t_j$ are the times associated to the observations $i$ and $j$, respectively. $\theta_1$ is the amplitude of the Gaussian Process (GP), $\theta_2$ is the decay timescale (i.e. the exponential timescale on which the modeled photometry departs from pure periodicity) or equivalently typical spots lifetime, $\theta_3$ is the period of the GP (expected to be close to $P_{\mathrm{rot}}$) and $\theta_4$ is the smoothing parameter that controls the amount of short-term variations that the fit can include (within a rotation cycle). For our purpose, we imposed a large decay timescale $\theta_2 = 300$ d to avoid the GP to fit rapidly evolving patterns. The resulting light curve is shown in Fig \ref{fig:Lc_tess}. We then modelled this filtered light curve with the same GPR but letting all 4 parameters free in order to derive an estimate of the typical timescale on which the main surface features evolve (found to be equal to $162^{+30}_{-25}$~d). This value is lower than that obtained by \cite{yu19} from ground-based $V$ measurements ($314^{+31}_{-29}$~d) most likely because TESS photometry is much more accurate and sensitive to small structures evolving rapidly. 
We also found $\theta_3 = 1.873\pm0.001$~d, consistent with the rotation period obtained by \cite{stelzer03} and the smoothing parameter $\theta_4=1.02\pm0.06$.
Considering only the data obtained before the end of SPIRou observations (2019 December 12), we computed median time and relative photometry every ten points, resulting in 757 photometric values, in order to get a smoother curve, to reduce computational time and to balance the relative weights of photometry and spectroscopy when applying ZDI (see Sec.~\ref{sec:brightness}). 

Despite the very-high-precision photometry provided by TESS, no periodic signal beyond that due to V410~Tau~A is detected in the light curve, which indicate that the temporal variations are attributable to the main star. In addition, the amplitude ratio between the peaks associated with component A and the noise in the TESS light curve periodogram (equal to $\unsim32$) provides an independant confirmation that the magnitude contrasts with the two other components in the $I_{\rm c}$ band is about 4 or more, in agreement with \cite{ghez97}.

\subsection{Additional photometric observations}
\label{sec:sec2.3}
\begin{figure}
    \centering
    \hspace*{-.4cm}
    \includegraphics[scale=0.32,trim={0.5cm 1.8cm 3cm 3cm},clip]{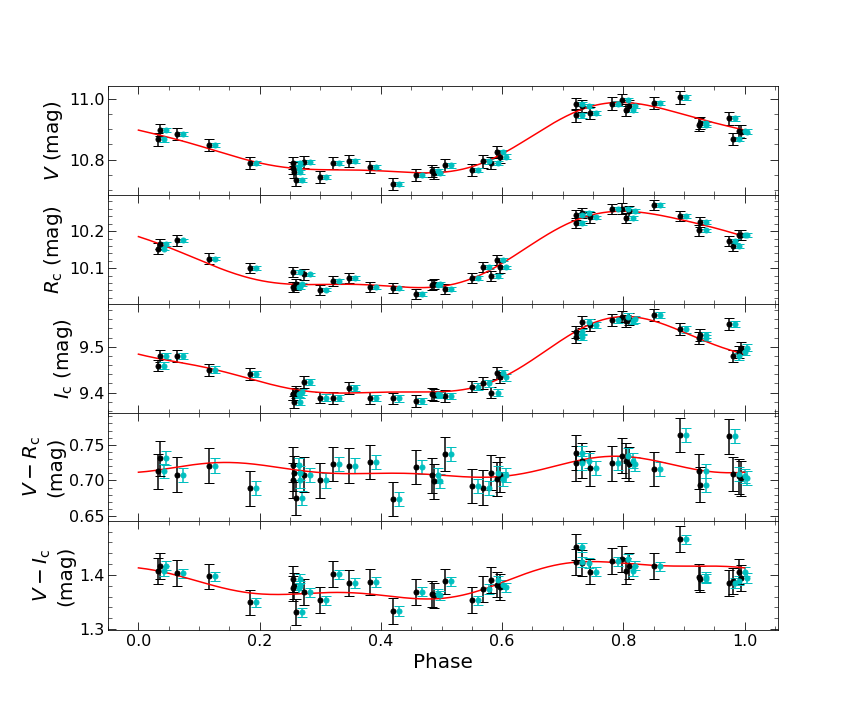}
    \caption{Ground-based photometric data in the $V$, $R_{\rm c}$ and $I_{\rm c}$ bands showing an empirical uncertainty of 20, 14 and 13 mmag, respectively (first to third panels), and $V-R_{\rm c}$ and $V-I_{\rm c}$ colour indexes (fourth and fifth panels). The cyan error bars correspond to the formal measurement uncertainties while the black error bars are the scaled-up empirical uncertainties (slightly shifted along the horizontal axis for display purposes). The red curves correspond to periodic fits to the $V$, $R_{\rm c}$ and $I_{\rm c}$ magnitudes including the fundamental frequency and the first two harmonics (three first panels). The red curves plotted on $V-R_{\rm c}$ and $V-I_{\rm c}$ were derived by subtracting the previous models. The light curve shows an amplitude of $0.231$~mag in the $V$ band, $0.212$~mag in the $R_{\rm c}$ band and $0.174$~mag in the $I_{\rm c}$ band. All light curves are phased with the same ephemeris as that used for SPIRou data (see Sec.~\ref{sec:sec2.1}).}
  
    \label{fig:compare_photometry}
\end{figure}

Multicolour photometry of V410~Tau was collected with the ground-based 1.25~m AZT-11 telescope at the CrAO from 2019 September 02 to December 18. Using a ProLine PL23042 CCD camera, 40 brightness measurements were collected in the $V$, $R_{\rm c}$ and $I_{\rm c}$ bands. For these estimates, the wTTS V1023~Tau was used as a comparison star as this star shows small variability amplitude. We note that V410~Tau has an average magnitude in the $V$ band equal to 10.85~mag (see Fig.~\ref{fig:compare_photometry}), thus consistent with observations of this star at previous epochs \citep{grankin08}. A full log of the CrAO photometric observations is given in Table~\ref{tab:log_crao}.  

We fitted the light curves in the $V$, $R_{\rm c}$ and $I_{\rm c}$ bands separately, using a periodic fit including the fundamental frequency and the first two harmonics.
As the formal uncertainties on the measured magnitudes (7~mmag for $V$ and $R_{\rm c}$, 5~mmag for $I_{\rm c}$) are underestimated, missing some sources of noise like the intrinsic variability of the observed star, we derived empirical estimates of these error bars by scaling them up to the values that ensure a unit reduced chi square for the fit in each band (using a sine wave plus 2 harmonics, see Fig.~\ref{fig:compare_photometry}). We find these empirical error bars to be equal to $20$, $14$, and $13$~mmag for $V$, $R_{\rm c}$ and $I_{\rm c}$, respectively. The models for $V-R_{\rm c}$ and $V-I_{\rm c}$ were then obtained by subtracting the models for the associated magnitudes while the error bars on these colours were estimated by propagating the uncertainties (Fig.~\ref{fig:compare_photometry}).

As expected we find that the light curve amplitude decreases with wavelength: $0.231\pm0.012$~mag in the $V$ band, $0.212\pm0.008$~mag in the $R_{\rm c}$ band and $0.174\pm0.008$~mag in the $I_{\rm c}$ band (with the error bars on the amplitudes derived from the empirical uncertainties on the measured magnitudes). As expected, TESS photometry (amplitude of $0.17656\pm0.0009$~mag) is consistent with that obtained from the ground in the $I_{\rm c}$ band. As all light curves are more or less consistent in shape, we will only use TESS data in the following as they are much more accurate than ground-based photometry.

\section{Stellar tomography}
\label{sec:sec3}
\subsection{Zeeman-Doppler Imaging}
\label{sec:sec3.1}
\begin{table*}

\caption{In this table we recall the data that were used to derive each of the main type of results outlined in this paper.}
\label{table:summary_ZDI}

\begin{tabular}{cccc}
\\
\hline 
\multicolumn{1}{c}{SPIRou Stokes~$I$} & \multicolumn{1}{c}{SPIRou Stokes~$I$ + TESS} & \multicolumn{1}{c}{SPIRou Stokes~$I$ + $V$} & \multicolumn{1}{c}{SPIRou Stokes~$I$} \\
\multicolumn{1}{c}{(M1: $\unsim$10 000 atomic + molecular lines)} & \multicolumn{1}{c}{(M1: $\unsim$10 000 atomic + molecular lines)} & \multicolumn{1}{c}{(M2: $\unsim$2000 atomic lines)} & \multicolumn{1}{c}{(M3: $\unsim$900 molecular lines)} \\ \hline
Brightness map & Brightness map & Brightness \& magnetic field maps & Brightness map \\
Differential Rotation (Stokes~$I$ only) & - & Differential Rotation (Stokes~$V$ only) & - \\
Radial velocities & Radial Velocities & Longitudinal field & - \\ \hline

\end{tabular}
\end{table*}

To recover maps of the surface brightness and/or of the magnetic field topology of the star, we applied ZDI \citep{semel89,brown91,donatibrown97,donati06} on our time series of Stokes~$I$ and/or Stokes $V$\footnote{Stokes $V$ profiles are obviously different from $V$ magnitudes despite sharing similar notations.  As there is no real ambiguities between the 2 quantities, we kept the notations unchanged.} LSD profiles. ZDI aims at constraining surface (brightness or magnetic) maps of rotating stars from time series of (Stokes $I$ and $V$) LSD profiles collected as the star rotates. To achieve this we proceed as follows. 
ZDI uses a conjugate gradient algorithm to deduce iteratively, from an initially non-spotted (resp. non-magnetic) distribution, the maps of relative brightness with respect to the quiet photosphere at SPIRou wavelengths (resp. of the magnetic field vector) at the surface of the star, until the corresponding synthetic LSD profiles fit the observed ones down to a unit reduced chi-square ($\chi^2_r$). This allows to look for the maximum-entropy solution of this optimization problem, i.e. the map containing the smallest amount of information capable of fitting the data down to the noise level.

In a second step, we applied ZDI to both SPIRou and TESS data simultaneously. We proceeded as in \cite{yu19} but this time including photometric data as part of the fit (instead of simply comparing the light curve predicted by ZDI with photometric observations). This is achieved by deriving, as part of the iterative imaging process and using Planck's law, the brightness contrast in the TESS bandpass that we expect from the one in the SPIRou bandpass (which we reconstruct through ZDI).  

In practice, we divide the surface of the star into a grid of a few thousand cells; to estimate the spectral contribution of each cell to the measured Stokes~$I$ and $V$ LSD profiles, we use Unno-Rachkovsky's analytical solution to the polarized radiative transfer equations in a Milne-Eddington atmosphere (see e.g. \citealt{landi04}) with appropriate values for relevant parameters such as the limb-darkening coefficient known to strongly depend on wavelength ($0.3\pm0.1$ in the $H$ band for $T_{\rm eff}=4500$~K and $\log g = 4$; \citealt{claret11}). The synthetic Stokes~$I$ (resp. Stokes~$V$) LSD profiles are then computed by integrating all the local Stokes~$I$ (resp. Stokes~$V$) LSD profiles over the visible stellar hemisphere, while the photometric values are computed by summing up the value of the continuum over all grid cells.

The poloidal and toroidal components of the magnetic field are decomposed into spherical harmonics \citep{donati06} while the photospheric relative brightness is computed independently for each cell of the grid.

In a first approach, we will assume that the star rotates as a solid body. We will then take into account differential rotation at the surface of V410~Tau in Sec.~\ref{sub:differential_rotation}.

In order to fit our LSD profiles, we chose a line model with mean wavelength, Doppler width and Landé factor of 1650~nm, 1.8~\kms\ and 1.2, respectively. As the depth of the LSD profiles varies depending on the mask and as we kept constant the Doppler width, our line models features an equivalent width (EW) of 1.5~\kms, 1.9~\kms\ and 1.4~\kms\ for the M1, M2 and M3 masks, respectively. 

Using ZDI on our Stokes~$I$ LSD profiles obtained from SPIRou data, we found $v\sin{i}~=~72.8~\pm~0.5$~ \kms\ for the line-of-sight-projected equatorial rotation velocity and $i=45\pm10^{\circ}$ for the inclination of the rotation axis to the line of sight. These values being consistent with those of \cite{yu19} within the error bars, we decided to follow \cite{yu19} and set $v\sin{i}=~73.2$ \kms\ and $i=50^{\circ}$. Using ZDI, we found that the bulk RV of the star is equal to $17.4\pm0.3$~\kms\ when considering M1, $18.1\pm0.3$~\kms\ when considering M2 and $16.0 \pm 0.6$ when considering M3. Although these values remain compatible within 3$\sigma$, we suspect that these differences may partly come from inaccuracies in the empirical mask line wavelengths, known to be less reliable for molecules than for atoms. It may also reflect a systematic RV blueshift of molecular lines with respect to atomic lines, that would suggest that atomic lines are more affected than molecular lines by the inhibition of convective blueshift by stellar magnetic activity for a reason yet to be clarified. 

Finally, as in \cite{yu19}, we set the maximum number of spherical harmonics to $\ell=15$ to describe the large-scale magnetic topology. 

A summary of the information provided by our ZDI analyses with the three different masks is provided in Table~\ref{table:summary_ZDI}.

\subsection{Brightness mapping}
\label{sec:brightness}

We focused on the Stokes~$I$ LSD profiles obtained with M1 to deduce the surface brightness map of the star. We performed two analyses, one with SPIRou data only and one considering simultaneously SPIRou and TESS data. 

\begin{figure}
    \centering \hspace*{-1cm}
    \includegraphics[scale=0.12]{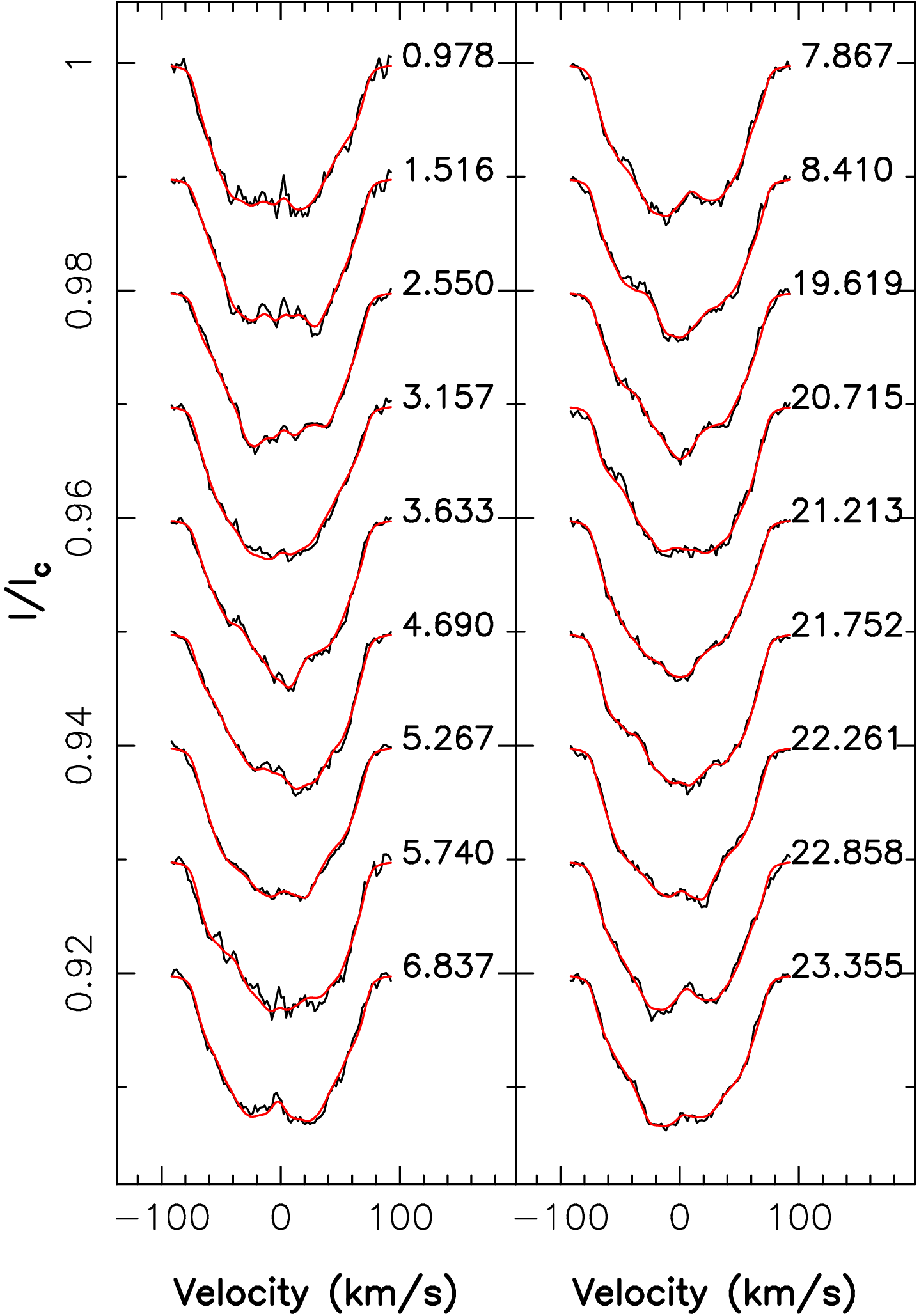}
    \caption{Stokes~$I$ LSD profiles obtained with the M1 mask (see Sec.~\ref{sec:sec2.1}). The observed profiles are shown in black while ZDI model (using SPIRou data only) is plotted in red. The rotation cycle associated with each profile is also mentioned on the right of each profile. Including the TESS data in ZDI does not lead to significant differences in the synthetic profiles.}
    \label{fig:LSD_profiles_I}
\end{figure}

\begin{figure}
    \centering
     \includegraphics[scale=0.08]{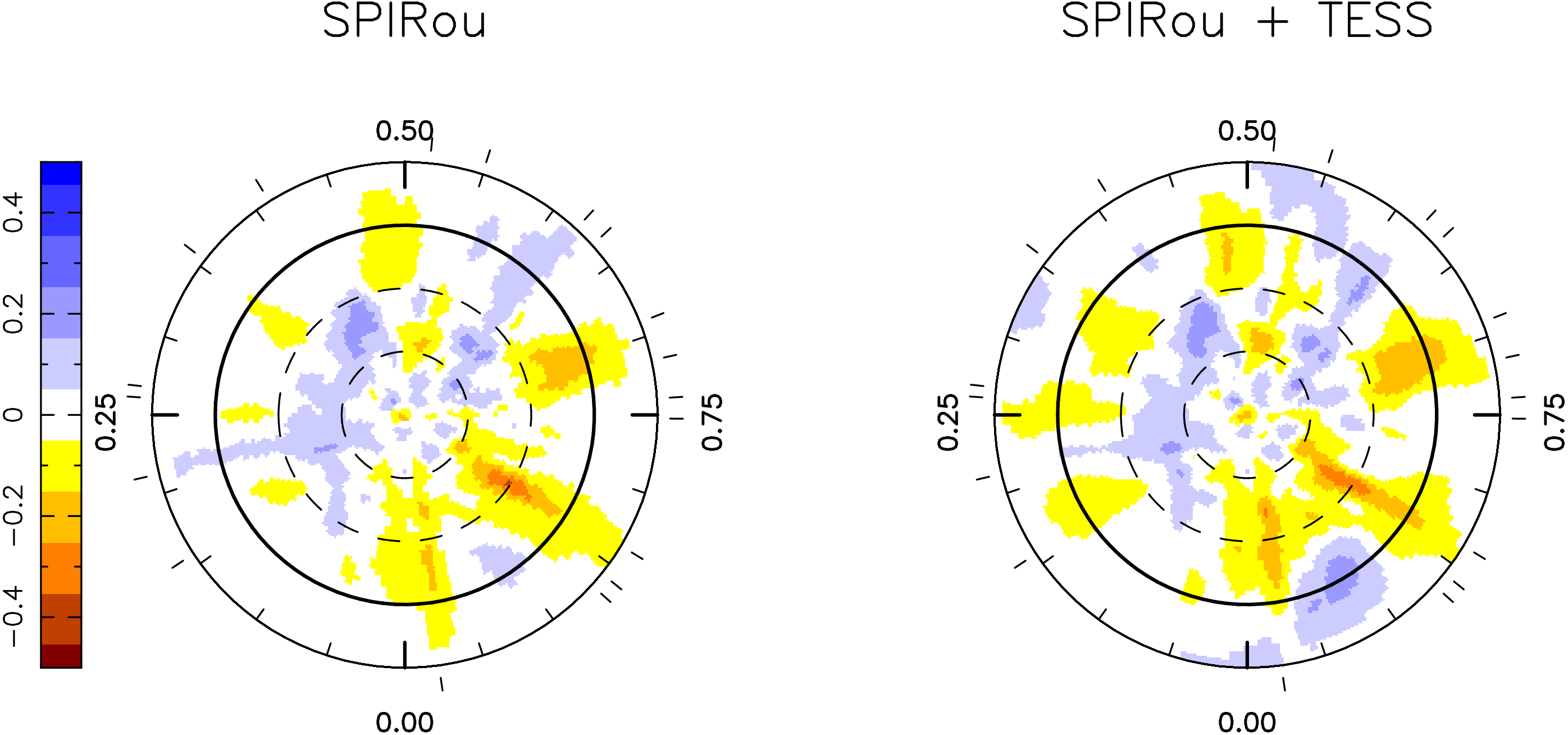}
    \caption{ZDI maps of the logarithmic relative surface brightness reconstructed from SPIRou data only (left) and from both SPIRou and TESS data (right). The maps are shown in a flattened polar view with the pole at the center, the equator represented as a bold black line and the 60$^\circ$ and 30$^\circ$ latitude parallels shown as dashed lines. The ticks around the star correspond to the phases of spectropolarimetric observations collected with SPIRou. Dark cool spots appear in yellow/red while the bright plages show up in blue.}
    \label{fig2}
\end{figure}

For the first analysis, we just applied ZDI on our time series of Stokes~$I$ LSD profiles that we fitted down to $\chi^2_r=1$ (see Fig.~\ref{fig:LSD_profiles_I}). The reconstructed surface brightness distribution is shown in the left panel of Fig. \ref{fig2}. We note the presence of large features, such as dark spots at phase 0.70 or 0.85, along with smaller ones. However, rather surprisingly, no polar spot is visible in this map from SPIRou data, whereas previous maps from optical data always showed conspicuous dark features covering the polar regions (e.g. \citealt{joncour94,hatzes95,yu19}).

\begin{figure}
    \centering 
     \includegraphics[scale=0.26]{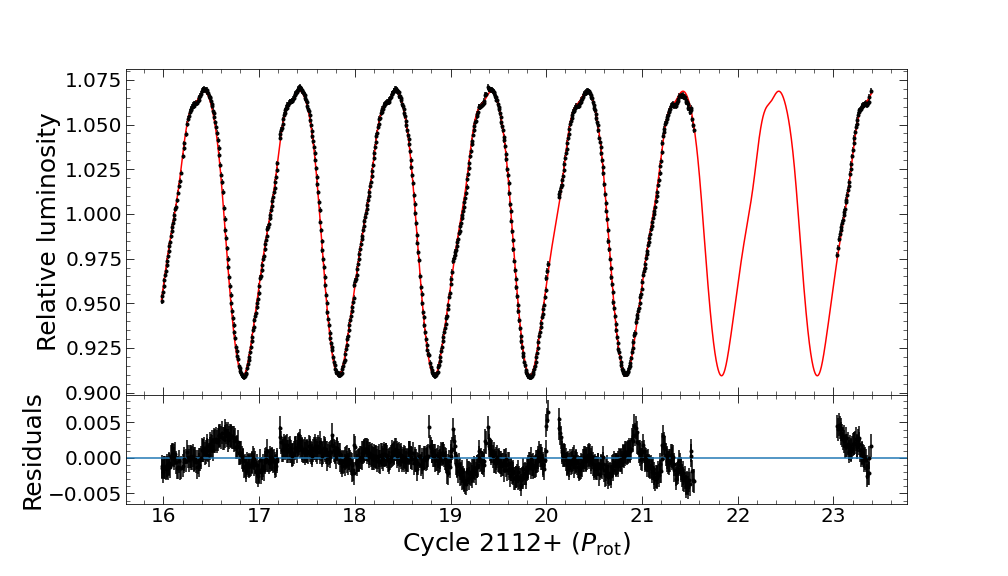}
    \caption{ZDI fit of the TESS light curve. \textit{Top panel:} the 757 relative photometry values from the TESS light curve are represented by the black dots (see Sec.~\ref{sec:sec2.2}). The fit of these points with ZDI is shown in solid red line. \textit{Bottom panel:} residuals exhibiting a dispersion of 1.6~mmag.}
    \label{fig:fit_tess}
\end{figure}

\begin{figure}
\centering
\includegraphics[scale=0.08]{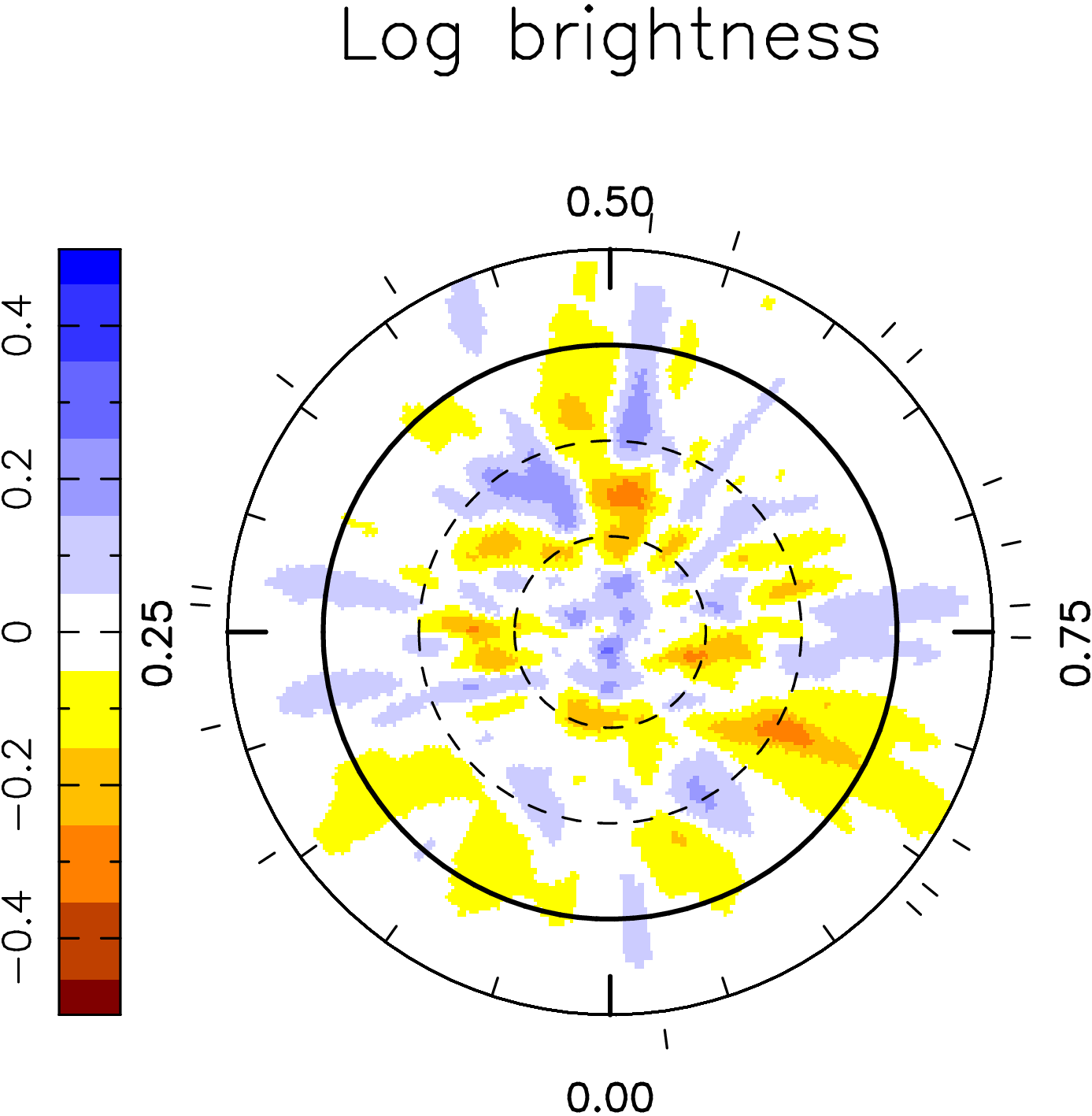}
\caption{Logaritmic relative surface brightness map obtained with ZDI by reconstructing the brightness with a mask containing only molecular lines. The star is shown in a flattened polar projection as in Fig. \ref{fig2}.}
\label{fig:mapQ_mol}
\end{figure}

\begin{figure}
    \centering
     \includegraphics[scale=0.08]{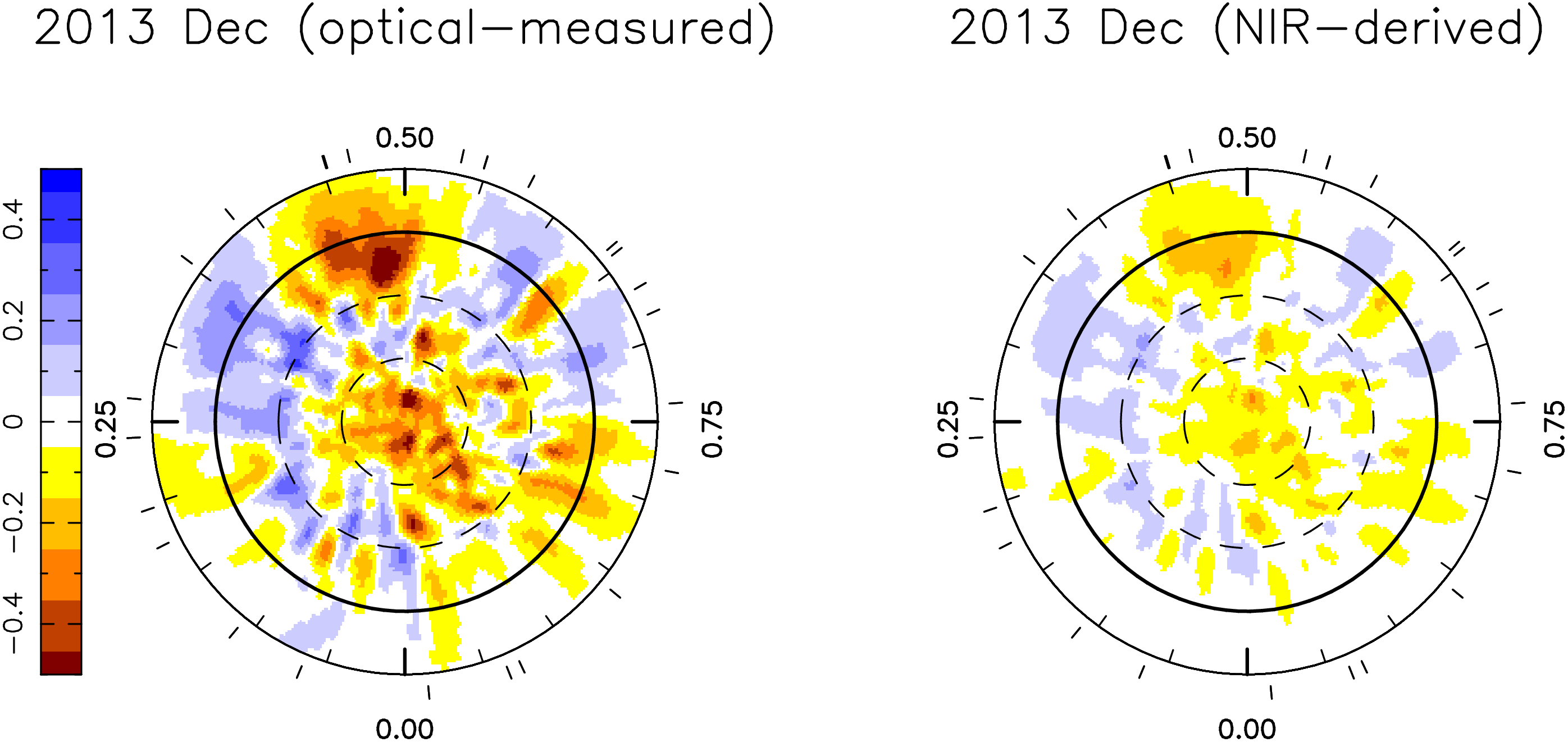}
    \caption{Logaritmic relative surface brightness maps obtained from 2013 optical data both in its original version (left panel, from \citealt{yu19}) and after rescaling to SPIRou wavelengths using Planck's law (right panel). The star is shown in a flattened polar view as in Fig. \ref{fig2}.}
    \label{fig:compare_2013_2019}
\end{figure}

\begin{figure}
    \centering \hspace*{-0.5cm}
    \includegraphics[scale=0.19]{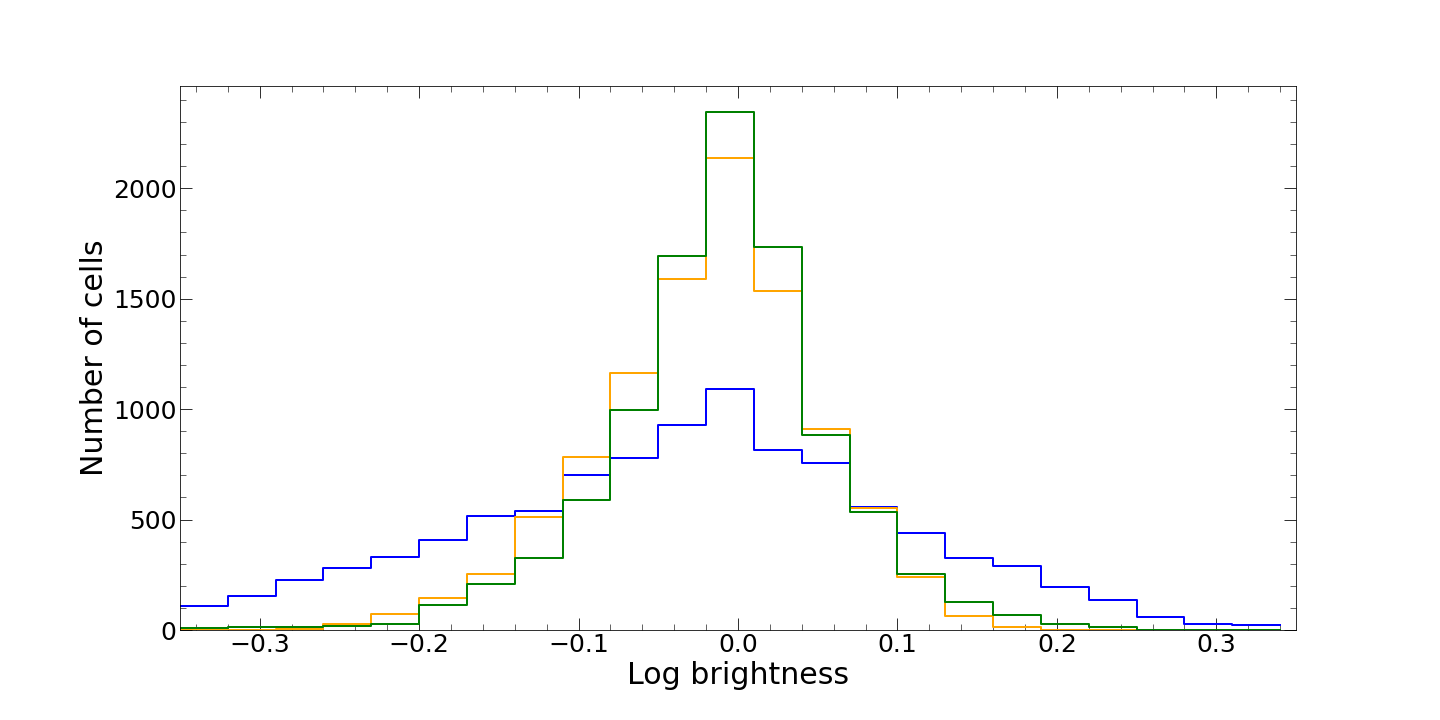}
    \caption{Histograms of contrasts for the maps shown in Fig. \ref{fig2} and \ref{fig:compare_2013_2019}. The blue distribution corresponds to the map obtained by \citet{yu19} with 2013 optical NARVAL data. The orange one corresponds to the 2013 optical map rescaled to SPIRou wavelengths using Planck's law, while the green one shows the distribution associated with the map obtained directly from our 2019 SPIRou data.}
    \label{fig:histograms}
\end{figure}

For our brightness reconstruction using both SPIRou and TESS data, we provided ZDI with the 757 photometric values (with error bars set to 1.6~mmag per data point) from the TESS light curve before 2019 December 12 (see Sec.~\ref{sec:sec2.2}) in addition to the Stokes~$I$ LSD profiles.  This yielded the reconstructed map and the fitted light curve shown in the right panel of Fig.~\ref{fig2} and in Fig.~\ref{fig:fit_tess}, respectively. Both spectroscopic and photometric data were fitted down to $\chi^2_r=~1$, with the fit of Stokes~$I$ profiles being almost identical to that obtained using SPIRou data only. We find that including photometry in addition to SPIRou data yields brightness maps with enhanced contrasts, especially at low latitudes, that were needed to fit the light curve at a RMS level of 1.6~mmag.

When fitting Stokes~$I$ profiles only, the spot coverage is close to 8.4\% while it increased to 10.7\% when TESS data were taken into account.This 2.3\% increase is significant, firstly because this difference is larger than the typical error we can expect on this parameter (of order 0.4\% in the context of this particular data set), and secondly since maximum entropy is profiled to provide the image containing the smallest amount of information. We come back on the origin of this difference in Sec.~\ref{sec:5.1}. We note that, in the fitted light curve, the residuals still exhibit correlated noise (at a level of 1.6~mmag RMS) that does not repeat from one rotation cycle to the next, and that likely reflects small-scale structures at the surface that evolve with time. (This explains a posteriori why the error bar on the TESS data points was set to this value).  

As no polar spot is visible whether we take into account the TESS data or not, we reconstructed the brightness distribution from Stokes~$I$ LSD profiles using the M3 mask, expected to be more sensitive to cooler regions (Fig.~\ref{fig:mapQ_mol} and \ref{fig:StokesI_M3}). Once again, the reconstructed map does not show a polar spot but displays some low-level differences with the map reconstructed with the M1 mask, e.g., the cool spot at phase 0.70 (see Fig~\ref{fig:mapQ_mol}).

Obviously, we do expect the brightness distribution to evolve over timescales of several years. However, we expect histograms of brightness contrasts to remain more or less the same in a given spectral range. In this context we can compare maps from NIR and optical data by rescaling, with Planck's law, brightness maps reconstructed from optical data (see left panel of Fig.~\ref{fig:compare_2013_2019} in the particular case of epoch 2013 December; from \citealt{yu19}) to the image one would have reconstructed in the NIR (right panel of Fig.~\ref{fig:compare_2013_2019}). We find that, as expected, the NIR map is less contrasted than the original optical map. Moreover, the histogram of rescaled optical data is comparable with that directly obtained from SPIRou data (Fig.~\ref{fig:histograms}).

\subsection{Magnetic analysis}
\label{sec:magnetic}

To reconstruct the large-scale magnetic topology, we focused on LSD profiles provided by mask M2, as it contains only lines with well known magnetic sensitivity. 

\begin{figure}
    \centering \hspace*{-0.5cm}
    \includegraphics[scale=0.12]{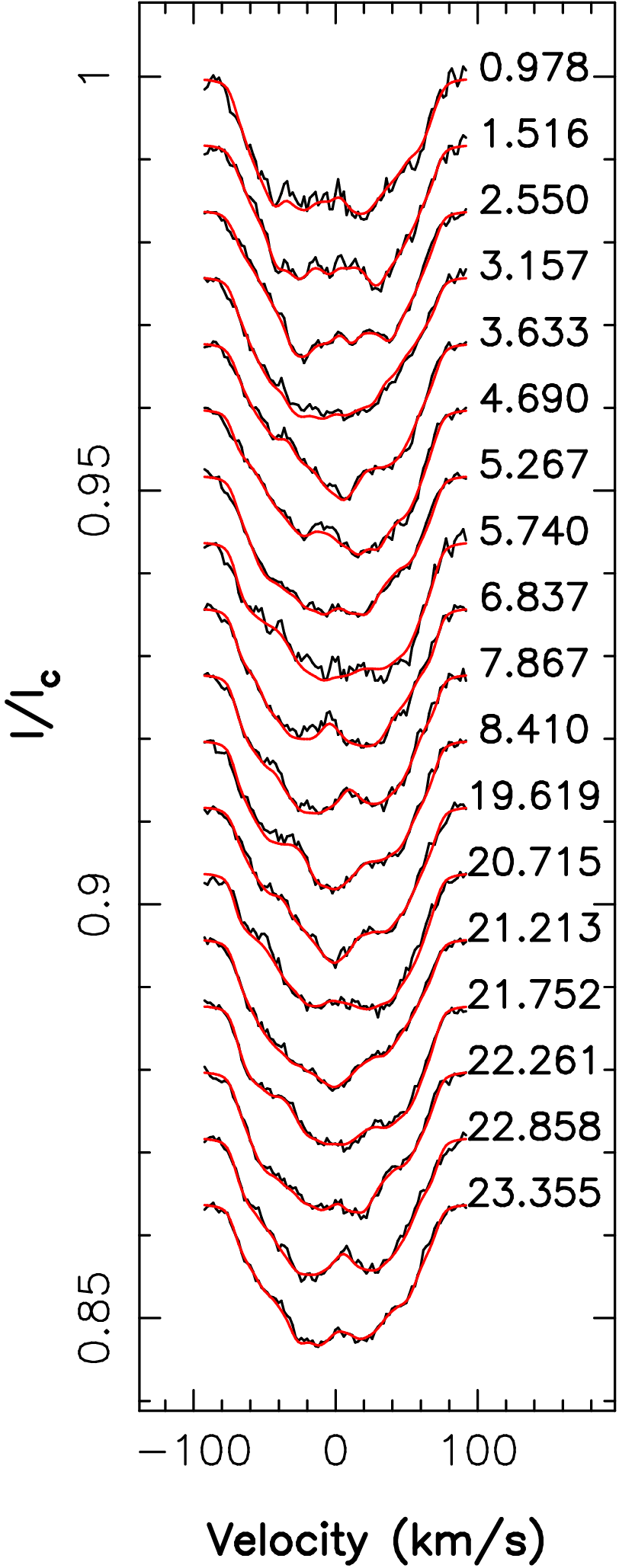} \hspace*{0.5cm}
    \includegraphics[scale=0.12]{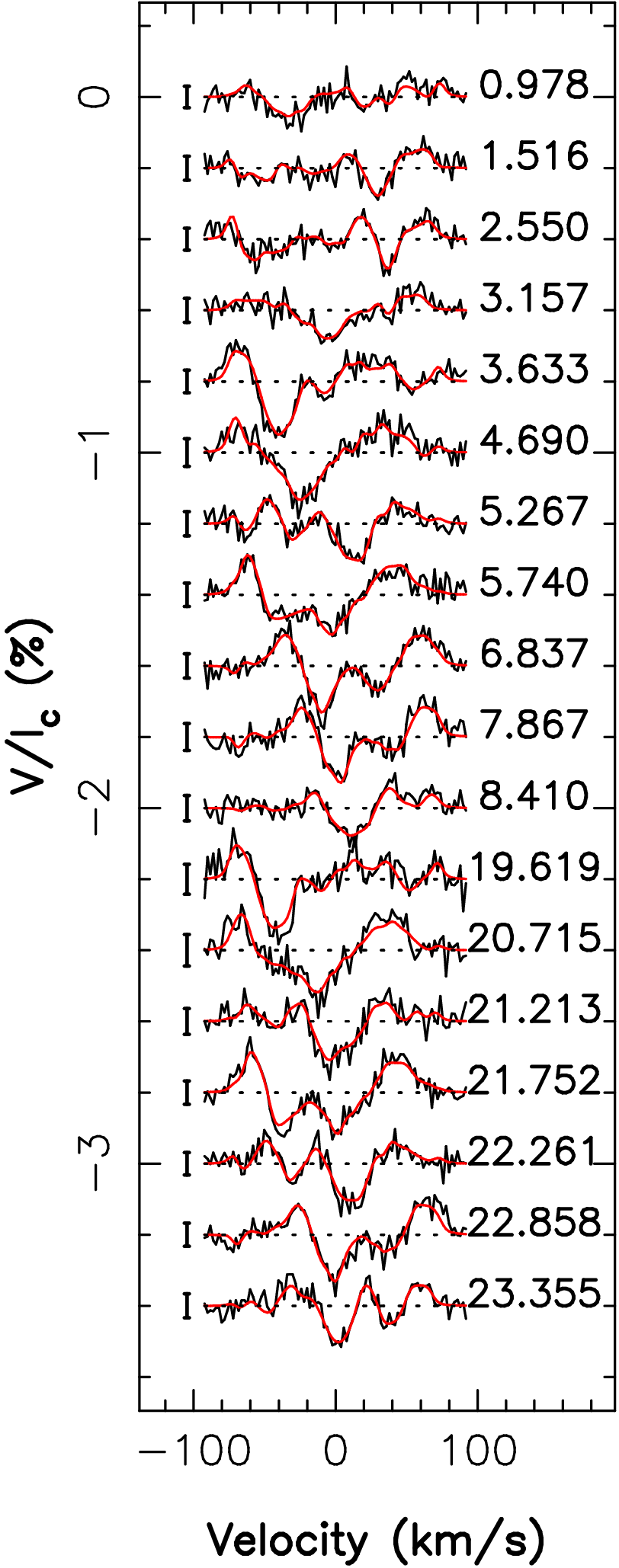}
    \caption{Stokes~$I$ (left) and Stokes~$V$ (right) LSD profiles obtained with a mask containing only atomic lines with well known Landé factors. The observed profiles are plotted in black while the ZDI fit is shown in red. The rotation cycle is mentioned on the right of each profile. 3$\sigma$ error bars are displayed on the left of each Stokes~$V$ profile. } 
   \label{fig:LSD_profiles_IV}
\end{figure}

\begin{figure*}
    \centering
    \includegraphics[scale=0.7]{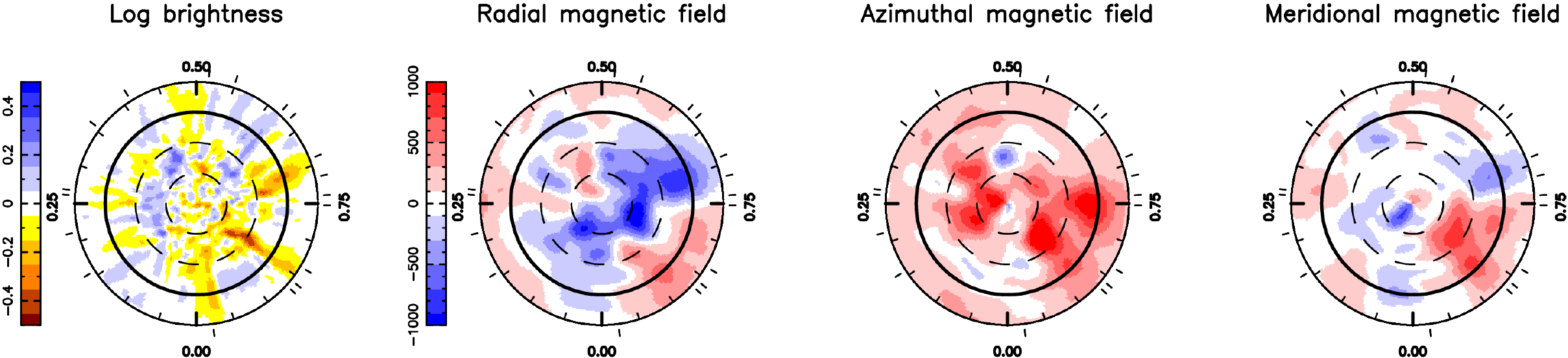}
    
    \caption{ZDI maps of the logarithmic relative surface brightness (1$^{\rm st}$ panel), and of the radial, azimuthal and meridional magnetic field components (2$^{\rm nd}$ to 4$^{\rm th}$ panels) obtained from a simultaneous fit of Stokes~$I$ and Stokes~$V$ LSD profiles. The description of the brightness map is as in Fig.~\ref{fig2}. For the magnetic field maps, red indicates positive radial, azimuthal and meridional fields that point outwards, counterclockwise and polewards, respectively. The star is shown in a flattened polar projection as in Fig. \ref{fig2}.}
    \label{fig:maps_Iv}
\end{figure*}

\begin{figure}
\centering \hspace*{-0.5cm}
\includegraphics[scale=0.27]{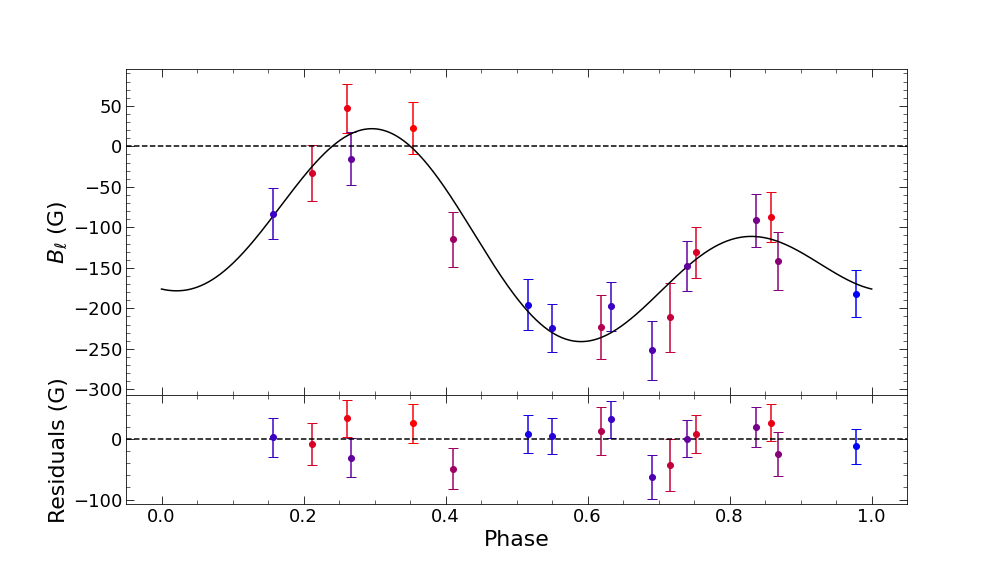}
\caption{Phase folded longitudinal field $B_\ell$. \textit{Top panel:} Observed values are represented by coloured dots, each color representing a different cycle. The black curve corresponds to the fit of our data with a sine curve with one harmonic. \textit{Bottom panel:} Residuals between the raw $B_\ell$ and the model, showing a dispersion of about 30~G. }
\label{fig:bl}
\end{figure}

We fitted simultaneously Stokes~$I$ and Stokes~$V$ LSD profiles with ZDI. Once again, we achieved a fit of the data down to $\chi^2_r =1$. The profiles and the associated maps are shown in Fig. \ref{fig:LSD_profiles_IV} and \ref{fig:maps_Iv}, respectively. We found a map of the logarithmic surface brightness similar to those reconstructed with mask M1 (with and without the TESS data). We see that the magnetic field is complex, with a topology similar to those derived by \cite{yu19}. We note that the magnetic field average strength is close to 410~G. The poloidal component of the field, which encloses about 60\% of the overall reconstructed magnetic energy, is essentially non-axisymmetric ($\unsim50\%$) and weakly dipolar ($\unsim35\%$) while the toroidal one presents the inverse properties ($\unsim70\%$ axisymmetric and $\unsim55\%$ dipolar). We find that the dipole component has a polar strength of 390~G and that its axis is tilted at 15$^\circ$ to the rotation axis, towards phase 0.70. 

We see no clear correlation between brightness and magnetic field maps obtained with ZDI as was already the case in maps derived from optical data \citep{yu19}.

We determined the longitudinal field\footnote{The longitudinal field is defined as the (algebraic) line-of-sight projected magnetic field component averaged over the visible hemisphere and weighted by brightness inhomogeneities.} $B_\ell$ at each epoch by computing the first moment of Stokes~$V$ profiles \citep{donati97}. The longitudinal field varies between about +50 and -250~G. The associated uncertainties range between 29~G and 42~G with a median of 32~G. We clearly see a periodic pattern in this index that can be fitted down to the noise level with a sine curve with one harmonic (period of $1.873\pm0.002$~d) as shown in Fig. \ref{fig:bl}.

\subsection{Differential rotation}
\label{sub:differential_rotation}

As our spectropolarimetric observations are spread over more than one month, the data can potentially exhibit some variability due to differential rotation. By computing the amount of shearing by latitudinal differential rotation that the surface (brightness or magnetic) maps experienced with time, ZDI allows one to estimate the surface differential rotation assuming a solar-like shear, given by:

\begin{equation}
    \Omega(\theta) = \Omega_{\mathrm{eq}} - (\cos \theta)^2 \mathrm{d}\Omega 
\end{equation}
where $\theta$ is the colatitude, $\Omega_{\mathrm{eq}}$ and $\mathrm{d}\Omega$ are the rotation rates at the equator and the difference of rotation rate between the pole and the equator, respectively. This differential rotation law was found to be successful at modeling the surface shear of low-mass stars, including those of rapidly rotating, fully convective dwarfs \citep{morin08}, including in particular V410~Tau \citep{yu19}.
One can measure both $\Omega_{\mathrm{eq}}$ and $\mathrm{d}\Omega$ by finding out the values that minimize the $\chi^2_r$ for a given amount of reconstructed information.

\begin{table}

   \centering
  \caption{Summary of differential rotation parameters of V410 Tau obtained thanks to ZDI. Column 1 indicates the parameters of interest. Column 2 and 3 refer to estimates provided by Stokes~$I$ (brightness reconstruction) and Stokes~$V$ profiles (magnetic reconstruction), respectively. In the first row, we report the number of points used into ZDI process. Rows 2-3 show the equatorial rotation rate $\Omega_\mathrm{eq}$ and the associated rotation period, along with their 68\% confidence interval. Rows 4-5 provide the pole-to-equatorial difference rate d$\Omega$ with its 68\% confidence interval and the rotation period at the pole. Row 6 gives the inverse slope of the ellipsoid in the $\Omega_\mathrm{eq}$-d$\Omega$ plane (also equal to $\cos^2\theta_s$, where $\theta_s$ is the colatitude of the gravity centre of the brightness or magnetic field distribution (see \citealt{donati2000})). Last rows give the rotation rate $\Omega_s$ at colatitude $\theta_s$ and the associated rotation period.}
  \label{tab:differential_rotation}
\begin{tabular}{lcc}
\hline
Parameter   & Stokes~$I$ data & Stokes~$V$ data  \\ \hline
n    & 1674  & 1674    \\
$\Omega_{\mathrm{eq}}$ (mrad d$^{-1}$) & $3358.8 \pm 0.5$  & $3358.7 \pm 0.4$  \\
$P_{\rm eq}$ (d) & $1.8707 \pm  0.0003$ & $1.8707 \pm  0.0002$ \\
d$\Omega$ (mrad d$^{-1}$)  & $6.4 \pm 2.2$ & $9.0 \pm 1.9$ \\
$P_{\rm pole}$ & $1.8742 \pm  0.0013$ & $1.8757 \pm  0.0011$ \\
$\cos^2 \theta_s$      & 0.241    & 0.220    \\
$\Omega_s$ (mrad d$^{-1}$)    & $3357.2 \pm 0.1$ &$3356.9 \pm 0.2$  \\
$P_{\rm s}$ & $1.87156 \pm  0.00006$ & $1.8717 \pm  0.0001$ \\

\hline
\end{tabular}
\end{table}

\begin{figure*}
    \hspace*{-.73cm}
    \centering 
    \begin{subfigure}[b]{0.49\textwidth}
         \centering
         \includegraphics[scale=0.35]{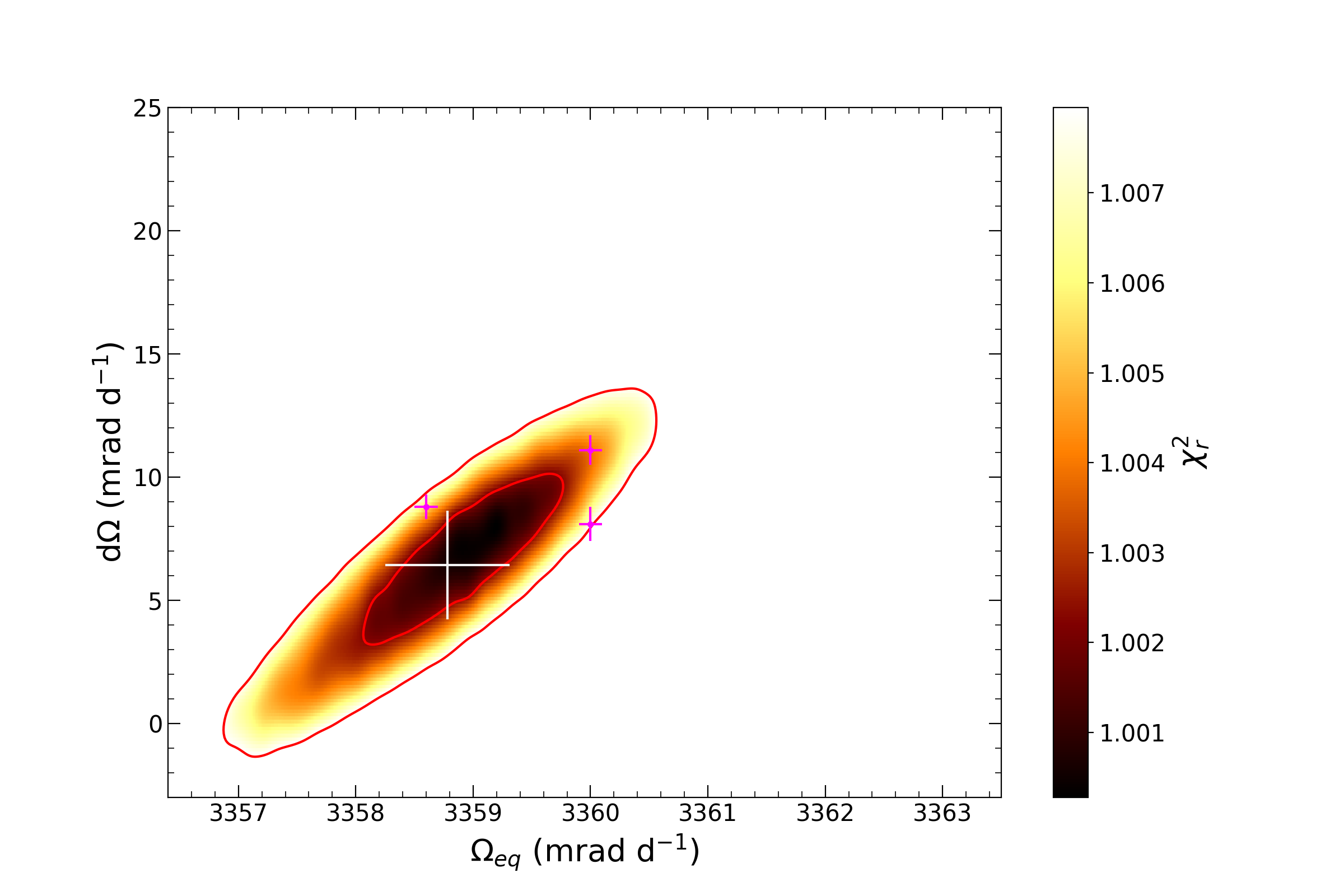}
         \caption{Stokes~$I$}
         \label{fig:subfig_contourI}
    \end{subfigure}
    \hfill
    \hspace*{-5cm}
    \begin{subfigure}[b]{0.52\textwidth}
         \centering
         \includegraphics[scale=0.35]{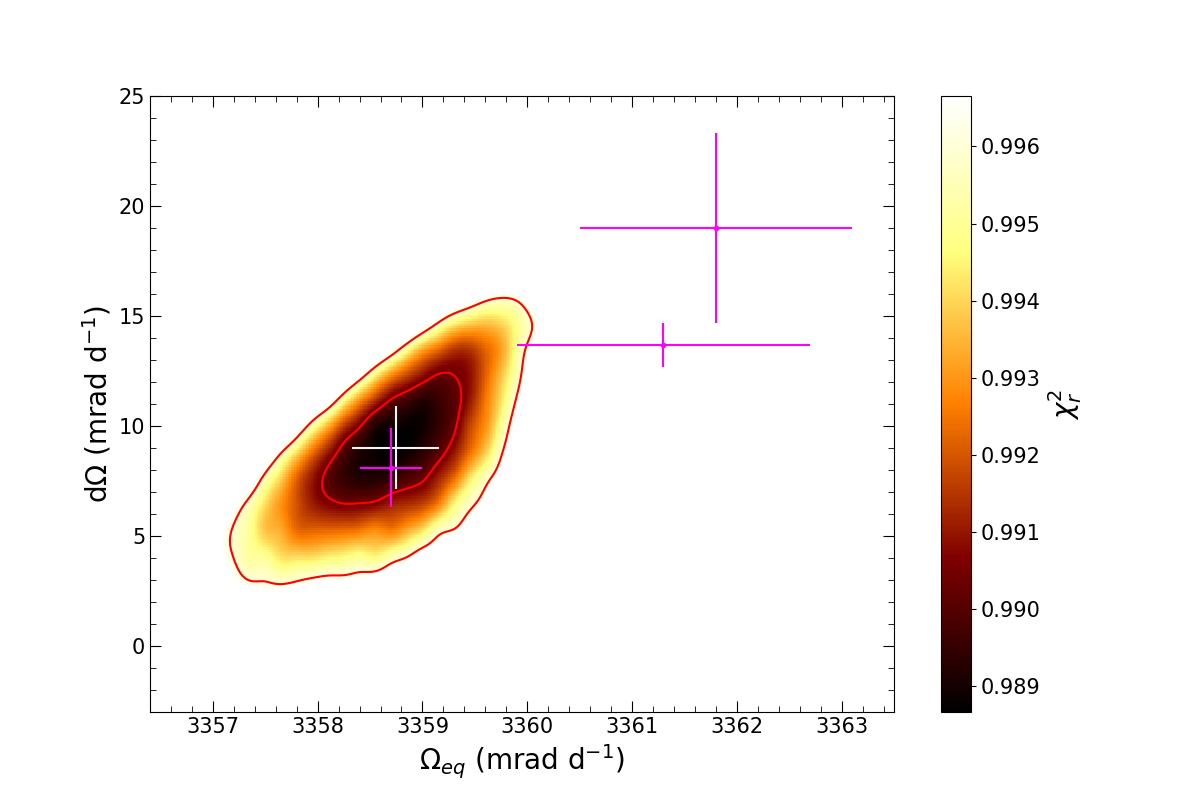}
         \caption{Stokes~$V$}
         \label{fig:subfig_contourV}
    \end{subfigure}
    \caption{Reduced $\chi^2$ map as a function of the differential rotation parameters $\Omega_{\mathrm{eq}}$, the equatorial rotation rate, and d$\Omega$, the pole-to-equator difference in rotation rate, obtained from (a) Stokes~$I$ and (b) Stokes~$V$ LSD profiles. The white cross indicates the optimal value with its associated error bars deduced from fitting a paraboloid to the \chisqr\ maps while the pink ones correspond to the estimates found by \citet{yu19}. Red ellipses define contours of 68\% (1$\sigma$) and 99.7\% (3$\sigma$) confidence levels for both parameters taken as pair. }
    
    \label{fig:contour_StokesI}
\end{figure*}

The $\chi^2_r$ maps derived from Stokes~$I$ and $V$ data respectively are shown in Fig. \ref{fig:contour_StokesI}, where contours of 68\% (1$\sigma$) and 99.7\% (3$\sigma$) confidence levels are depicted.
We find from Stokes~$I$ LSD profiles that $\Omega_{\mathrm{eq}} = 3358.8 \pm 0.5$~mrad~d$^{-1}$ and $\mathrm{d}\Omega = 6.4 \pm 2.2$~mrad~d$^{-1}$, while Stokes~$V$ LSD profiles yield $\Omega_{\mathrm{eq}} = 3358.7 \pm 0.4$~mrad~d$^{-1}$ and $\mathrm{d}\Omega = 9.0 \pm 1.9$~mrad~d$^{-1}$, both estimates being mutually consistent within 1.5~$\sigma$. This implies that the rotation period ranges from about 1.8707~d ($\pm 0.0003$ from Stokes~$I$ and $\pm 0.0002$ from Stokes~$V$) at the equator to $1.8742 \pm 0.0013$~d (from Stokes~$I$) or $1.8757\pm0.0011$~d (from Stokes~$V$) at the pole. Compared to previous shear detections in the optical \citep{yu19}, we find that the uncertainty on $\Omega_{\rm eq}$ and d$\Omega$ obtained from Stokes~$I$ are about 5 and 3 times larger, respectively, which is likely related to the lower amount of data in our set as well as to the lower contrast of the reconstructed brightness features. On the contrary, uncertainties derived from Stokes~$V$ profiles are comparable to that estimated from optical data at previous epochs.

We also computed the colatitude corresponding to the barycentre of the brightness and magnetic field distributions from the slope of the major axis of the confidence ellipse. In particular, our estimate from Stokes~$I$ LSD profiles is slightly larger than the ones derived by \cite{yu19} in the optical at previous epochs, suggesting that large surface features are indeed migrating poleward as speculated by these authors.

Table \ref{tab:differential_rotation} gathers our results about differential rotation.

\section{Stellar activity}
\label{sec:sec4}
\subsection{Radial velocities}
\label{sec:sec4.1}
\begin{figure*}
	\centering \hspace*{-1cm}
	\includegraphics[scale=0.4,trim={1.5cm 1.cm 3cm 2cm},clip]{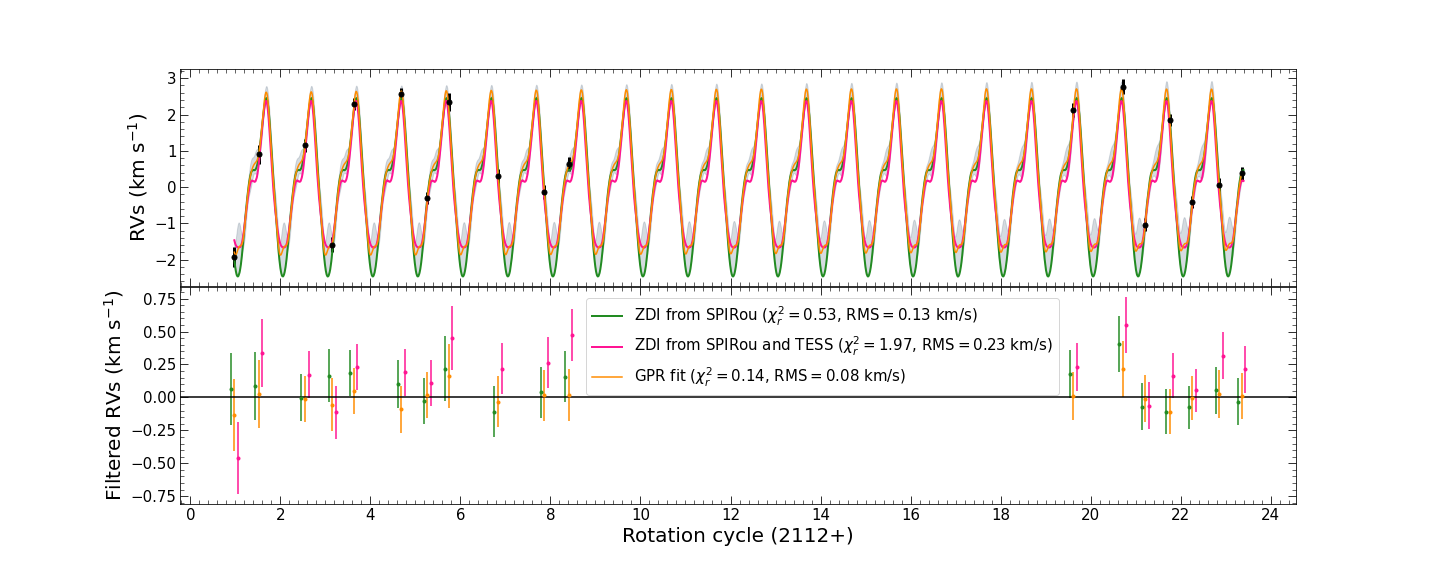}
	\caption{RVs of V410 Tau in 2019 November and December. \textit{Top panel:} Raw observed RVs have a dispersion of 1.40 \kms\ and are represented by black dots with their associated error bars. The green and pink curves are ZDI models obtained from brightness reconstruction considering SPIRou data only or SPIRou and TESS data simultaneously, respectively. The orange curve corresponds to the Gaussian process regression with its associated 1$\sigma$ confidence area in light grey.  \textit{Bottom panel:} Filtered RVs for each of the three models. The color code of the points is the same as for the curves in the top panel. The dispersion of the filtered RVs is 0.13, 0.23 and 0.08 \kms\ for green, pink and orange data, respectively. Filtered RVs of different colors at each observation phase are slightly shifted along the horizontal axis for graphics purposes.}
	\label{fig:RV}
\end{figure*}

We computed the RV of V410 Tau at each observed phase as the first moment of Stokes~$I$ LSD profiles \citep{donati17} for both observed (obtained with M1) and ZDI synthetic ones. With the set of synthetic profiles, we simulated noisy profiles with the same SNRs as the observed ones. For several realisations of the noise we computed the RV and then the dispersion of these measurements to estimate the error bars on our actual radial velocity data. We compared raw observed RVs with synthetic RV curves obtained from the maps in Fig. \ref{fig2} and we computed activity filtered RVs as the difference between the observations and the model (Fig.~\ref{fig:RV}). We note that both maps lead to models that fit reasonably well the data.

In a second step, we used GPR to model the impact of stellar activity on the observed RVs. For that, we chose the same kernel as that presented in Sec.~\ref{sec:sec2.2}, in Eq. \eqref{eq1}. Moreover, we added an additional term representing a potential excess of uncorrelated noise (in case our RV error bars are underestimated). The log likelihood function $\log \mathcal{L}$ we wanted to maximize becomes:

\begin{equation}
	\log \mathcal{L} = -\frac{1}{2} \left( N\log2\pi + \log |\mathrm{K} + \mathrm{\Sigma} + \mathrm{S} | + y^T( \mathrm{K} + \mathrm{\Sigma} + \mathrm{S})^{-1}y \right)
\end{equation}

where $\mathrm{K}$, $ \mathrm{\Sigma}$ and $\mathrm{S}$ denote the covariance matrix with a quasi-periodic kernel, the diagonal matrix containing the variance of the observed RVs, and the diagonal matrix containing the additional noise $s$ to the square. $N$ refers to the number of points (i.e. number of observed RVs) and $y$ corresponds to the observed raw RVs. 

\begin{figure}
	\includegraphics[scale=0.4]{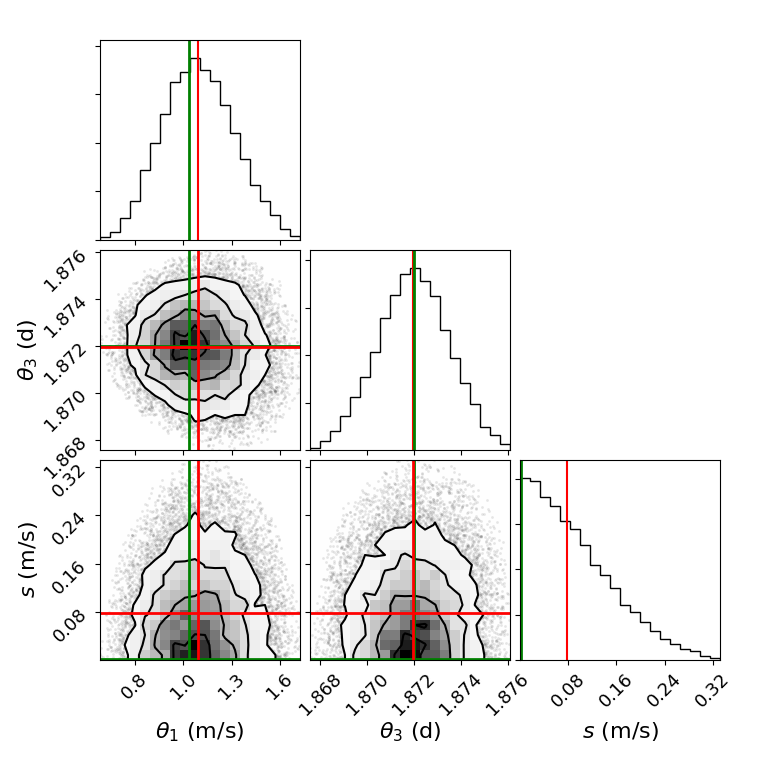}
	\caption{Phase plot of the posterior distribution of the three parameters let free, returned by the MCMC sampling. The best value for each parameter is chosen as the median value of the posterior distributions distributions shown in red line. We found $\ln\theta_1 = 0.08\pm0.18$, $\theta_3 = 1.872\pm0.002$ d and $s = 0.08\pm0.07$~\kms. We also traced the values that maximizes the posterior distributions in green lines. The plot has been done with the \texttt{CORNER PYTHON} module \citep{corner}.}
	\label{fig:corner_plot_RV}
\end{figure}

Given the low number of RV points, we chose to fix 2 of the 4 hyperparameters, namely the decay timescale $\theta_2$ (exponential timescale on which modeled RVs depart from pure periodicity) and the smoothing parameter $\theta_4$ (controling the amount of short-term variations in the fit) at 160~d (as derived from TESS light curve) and 0.35, respectively, following \cite{yu19}\footnote{As shown in \cite{klein20}, surface features induce more complex modulation on RVs than on light curves, hence the smaller value of the smoothing parameter $\theta_4$ used here compared to that derived when fitting the TESS data (see Sec.~\ref{sec:sec2.2}).}. 
Through a Markov Chain Monte Carlo (MCMC) approach using the \texttt{EMCEE PYTHON} module \citep{emcee}, we sampled the posterior distribution of the other parameters given the priors listed in Table \ref{tab:prior_RV}. We ran our MCMC on 5000 iterations of 100 walkers, and removed a burn-in period of 250~iterations, that is about 5 times larger than the autocorrelation time of the chain ($\unsim50$~iterations). We then chose the median of these posterior distributions as best values for the free parameters.
The resulting phase plot is illustrated in Fig. \ref{fig:corner_plot_RV}. The amount of excess uncorrelated noise in the data (modeled with $s$) is found to be compatible with 0. From the best set of parameters, we obtained the GPR fit that is shown in Fig. \ref{fig:RV}. 

Each model yields a synthetic RV curve that we compared to our RV measurements; the corresponding \chisqr\ values are equal to 0.53 when applying ZDI to SPIRou data alone, 1.97 when ZDI is applied to combined SPIRou and TESS data and 0.14 when applying GPR.
The corresponding dispersion of the activity-filtered RVs is about twice lower when we use GPR (0.08~\kms) rather than ZDI (0.13~\kms\ and 0.23~\kms, using SPIRou data alone or both SPIRou and TESS data). We come back on the potential origin of this difference in Sec.~\ref{sec:sec5.5}. 
We note that these dispersions are consistent with the typical uncertainty on our RV measurements ($\unsim180$~\ms), demonstrating that our models are successful at reproducing the activity-induced RV variations, both for atomic and molecular lines (and despite the difference in bulk RVs for both sets of lines).

Applying the FF' method \citep{aigrain12} to the light curve predicted with ZDI, we can investigate the precision level at which this technique can mitigate activity in RV curves. We find that RV residuals exhibit a dispersion of 810~\ms\ RMS, i.e., 6-7$\times$ larger than those predicted with ZDI (130~\ms\ RMS), confirming that the FF' technique is not adequate for filtering out RV curves of moderately to rapidly rotating active stars, whose brightness distributions are often rather complex.

\begin{table}
	
	\caption{Priors used for the MCMC sampling for the GPR on raw RVs and median values of the hyperparameters posterior distributions. For the uniform priors, we give the lower and upper boundaries of the interval while for the modified Jeffreys prior \citep{gregory07} we give the knee value. For $\theta_2$ and $\theta_4$ we mention the value we imposed. }
	\label{tab:prior_RV}
	\centering

	\begin{tabular}{lcc}
	\hline
	
	Hyperparameter & Prior & Estimate \\ \hline
	
	$\ln\theta_1$ [$\ln(\mathrm{km\,s^{-1}})$] & Uniform (-10, 10) & $0.08\pm0.18$ \\
	$\theta_2$ [d] & 160  & \\
	$\theta_3$ [d] & Uniform (0.9 $P_{\rm rot}$, 1.1 $P_{\rm rot}$) & $1.872\pm0.002$ \\
	$\theta_4$  & 0.35 & \\
	$s$ [$\mathrm{km\,s^{-1}}$] & Modified Jeffreys ($\sigma_{\rm RV}$) &$0.08\pm0.07$  \\ 
	& & (compatible with zero)\\ \hline
	\end{tabular}
	
\end{table}

\subsection{Activity indicators}
\label{sec:activity_index}

We focused on three main activity indicators: the \ion{He}{i} triplet at 1083~nm, the Paschen $\beta$ (Pa$\beta$) and Brackett $\gamma$ (Br$\gamma$) lines at 1282~nm and 2165~nm, respectively (Fig.~\ref{fig:spectral_lines}).

We identified a flare on December 09 (cycle 21.752) with the corresponding spectrum being blueshifted by 100~\kms\ with respect to the stellar rest frame, and the flux in the three lines being stronger than the typical one, especially in the Pa$\beta$ line (Fig.~\ref{fig:spectral_lines}). This flare occurs in the main gap of the TESS data (BJD 2,458,826.5 to BJD 2,458,829.3, see Fig.~\ref{fig:Lc_tess}) and therefore does not show up in the light curve. In addition, two observations collected just after the flare (i.e. on December 10 and 11) were also affected by the flaring episode and we thus decided not to use these three observations for our analyses based on the \ion{He}{i} triplet and Pa$\beta$ and Br$\gamma$ lines. We also identified a feature in the red wing of Pa$\beta$ (at $+120$~\kms), likely tracing Ti, Ca and Fe lines blending with Pa$\beta$, that does not vary more than the continuum around this line and is thereby expected not to affect significantly our analyses. We then compute the amount by which the equivalent width of these lines vary as a result of activity, which we call activity `equivalent width variations' (EWVs). In the stellar rest frame, we divided each telluric-corrected Stokes~$I$ spectrum by the median spectrum shown in Fig.~\ref{fig:median_lines}, yielding the median-divided spectra in Fig.~\ref{fig:spectral_lines_residuals}. The activity EWVs are then defined as the EW of these median-divided spectra counted as negative when absorption is larger than average. The values of the EWVs (and the corresponding error bars) were obtained through a gaussian fit to the median-divided spectra, using a gaussian of full-width-at-half maximum equal to 130~\kms\ centred on the stellar rest frame (consistent with the median \ion{He}{i}, Pa$\beta$ and Br$\gamma$ median profiles). We note that an activity indicator equal to 0 at a specific epoch does not indicate a lack of detection but rather that the corresponding profile is identical to the median one. The EWVs are provided in Table~\ref{table1}.

For each line, we assumed equal error bars for all spectral points of all observations, which we set to the dispersion between spectra in the continuum about each line (equal to 0.014 for \ion{He}{i} and Br$\gamma$, and 0.008 for Pa$\beta$, and tracing mostly photon noise). The corresponding error bars we derive for the EWVs are equal to 2, 1 and 4~pm, respectively. As the integrated flux in the \ion{He}{i} and Pa$\beta$ lines is variable at a higher level than that expected from photon noise, we empirically derived the error bars on the EWVs of the three lines using the same method as in Sec.~\ref{sec:sec2.3} (likely overestimating the uncertainties) to account for the intrinsic variability and other main sources of noise that cannot be easily quantified. We achieved this by fitting the \ion{He}{i} and Pa$\beta$ EWVs with a sine curve (including 2 harmonics for \ion{He}{i}), whereas the Br$\gamma$ EWVs (showing essentially no variation with time) were fitted with a constant, yielding error bars of 25, 3 and 6~pm for the \ion{He}{i}, Pa$\beta$ and Br$\gamma$ lines, respectively. 
To assess the significance of our models, we computed the \chisqr\ when fitting a constant instead of a periodic curve for the \ion{He}{i} (\chisqr$=13.6$) and Pa$\beta$ (\chisqr$=3.5$) lines, yielding probabilities of 0 and $3.6\,10^{-6}$, respectively, for the detected modulation to be spurious by chance. Even with these simple models and pessimistic estimates of the error bars, we detected a significant modulation of both the \ion{He}{i} and Pa$\beta$ EWVs. We however caution that the false alarm probabilies (FAPs) we quote, assuming white noise, may be underestimated if correlated noise dominates, even though pessimistic error bars were used.

The activity EWVs reveal enhanced absorption in phase range 0.4-0.6 for both the \ion{He}{i} triplet and the Pa$\beta$ line. This feature is also seen in the dynamic spectrum of the \ion{He}{i} triplet (Fig.~\ref{fig:dynamic_spectra}).

\begin{figure}
    \centering
    \begin{subfigure}{0.49\textwidth}
         \centering
         \includegraphics[scale=0.285]{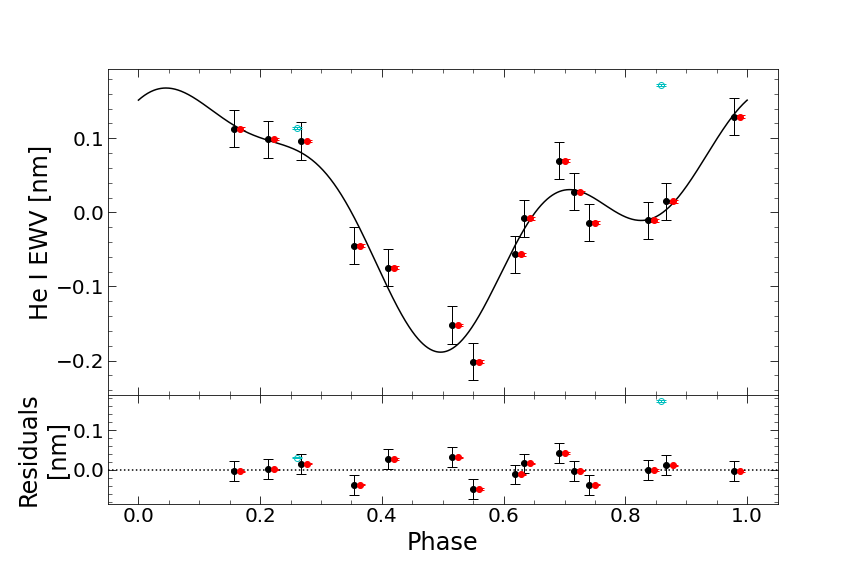}
         
    \end{subfigure}
    \hfill
    \begin{subfigure}{0.49\textwidth}
         \centering
         \includegraphics[scale=0.285]{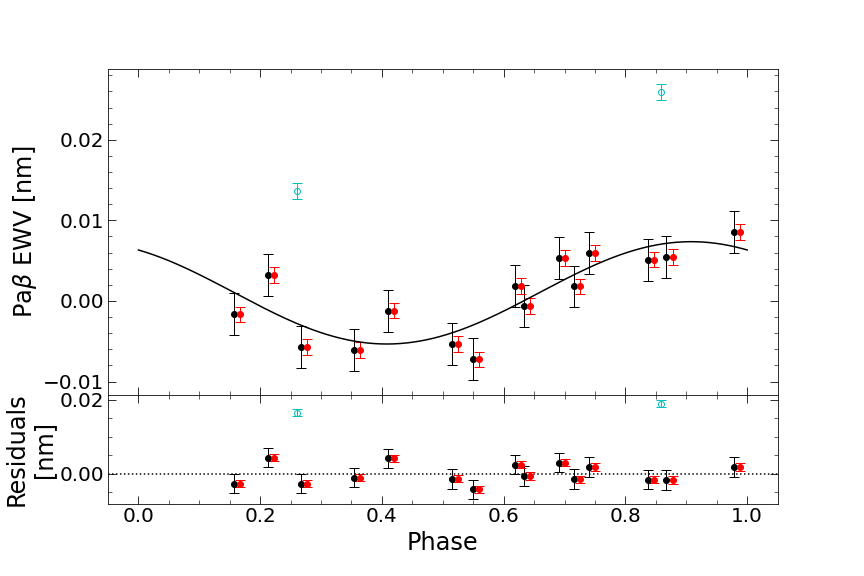}
         
    \end{subfigure}
    \hfill
    \begin{subfigure}{0.49\textwidth}
         \centering
         \includegraphics[scale=0.285]{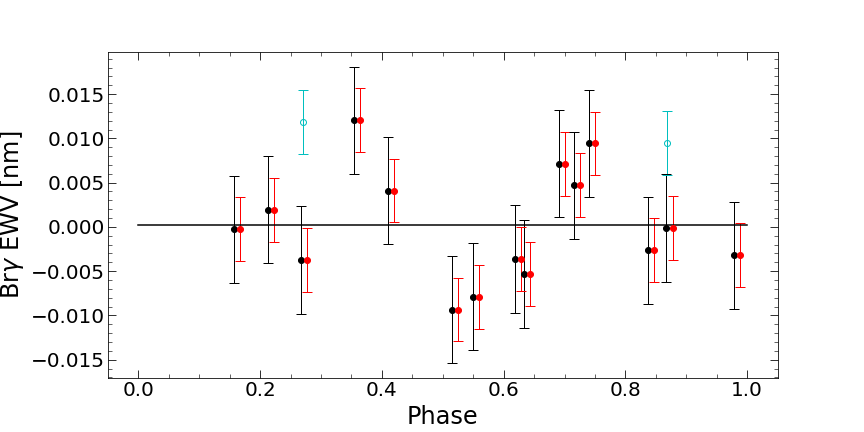}
         
    \end{subfigure}
    
    \caption{Phase folded activity EWVs derived from the \ion{He}{i} triplet at 1083.3~nm (first panel), Pa$\beta$ (second panel) and Br$\gamma$ lines (third panel) after removing the profile affected by the main flare at cycle 21.752. For the first two panels, the fit to the EWVs is shown in black line in the top panel while the bottom panel displayed the residuals between the EWVs and the best fit. In all panels, the red error bars correspond to those expected from the dispersion between spectra in the adjacent continuum (tracing photon noise) while the black ones were set to ensure a unit \chisqr\ fit to the data (thereby tracing intrinsic variability as well, and equal to 25, 3 and 6~pm for the \ion{He}{i}, Pa$\beta$ and Br$\gamma$ EWVs, respectively). The cyan open circles and error bars represent the two observations collected after the flare (not taken into account for the fit). The red error bars are slightly shifted along the horizontal axis for clarity purposes.}

    \label{fig:proxies_sinus}
\end{figure}

\subsection{Correlation matrices}
\label{sec:sec4.3}
From the median-divided spectra, we computed autocorrelation matrices for each of the three lines, considering the line relative intensities within an interval of $\pm200$~\kms. The coefficient $C_{ij}$ between velocity bins $i$ and $j$ is defined as:

\begin{equation}
   C_{ij} = r_{ij} \sqrt{\sigma_i \sigma_j}
   \label{eq:4}
\end{equation}

where $r_{ij}$ is the Pearson linear coefficient between the two velocity bins, $\sigma_i$ and $\sigma_j$ are the standard deviation in the velocity bins $i$ and $j$, respectively. This definition of the unnormalised coefficient allows us to estimate the relative importance of the correlations, as a high value of $r_{ij}$ associated with a high level of variability is better emphasized than a high value of $r_{ij}$ associated with a low level of variability.

\begin{figure*}
	\centering
	\begin{subfigure}[b]{0.3\textwidth}
         \centering
         \includegraphics[scale=0.22, trim= 0cm 0cm 2cm 1 cm, clip]{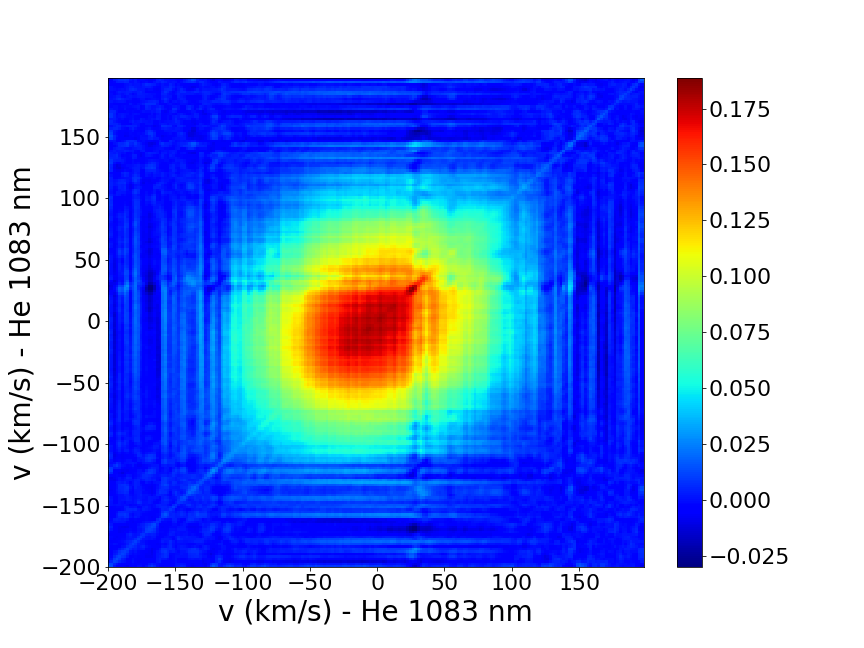}
         
    \end{subfigure}
    \hfill
    \begin{subfigure}[b]{0.3\textwidth}
         \centering
         \includegraphics[scale=0.22, trim= 0cm 0cm 2cm 1 cm, clip]{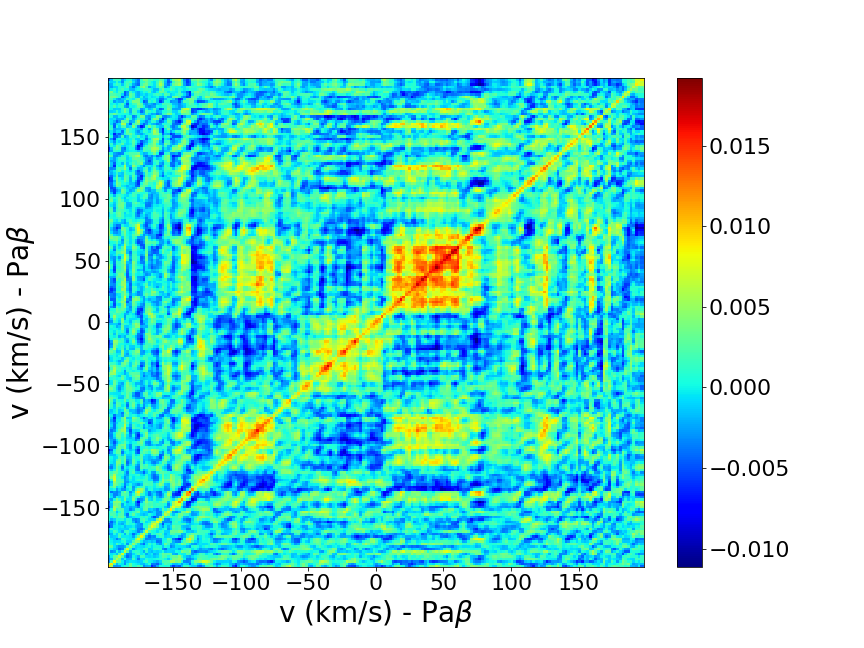}
         
    \end{subfigure}
    \hfill
    \begin{subfigure}[b]{0.3\textwidth}
         \centering
         \includegraphics[scale=0.22, trim= 0cm 0cm 2cm 1 cm, clip]{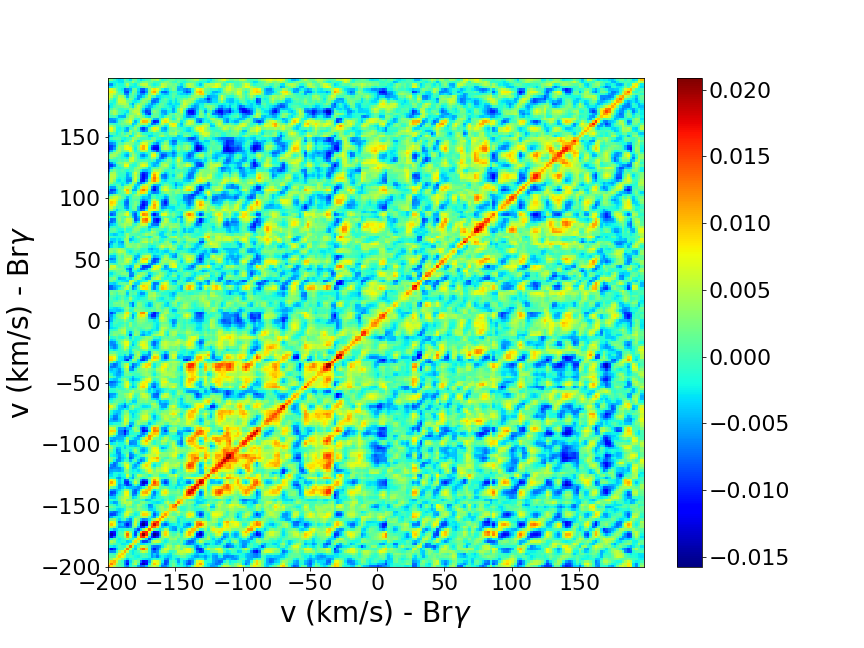}
        
    \end{subfigure}
    
	\caption{Autocorrelation matrices for \ion{He}{i} triplet (left panel), Pa$\beta$ (middle panel) and Br$\gamma$ (right panel) after removing the three observations affected by a flaring episode. The colorbars refer to the value of the coefficients as defined by Eq.~\eqref{eq:4}, with important correlation represented by reddish colours. The color scale depends on the level of variability for the considered line which is much larger for the \ion{He}{i} triplet than for the two other lines.}
	\label{fig:autocorrelation}
\end{figure*}

The autocorrelation matrix of the \ion{He}{i} triplet reveals that the entire profile correlates well with itself (left panel of Fig.~\ref{fig:autocorrelation}). As a strong correlation indicates a common origin for the components, the observed correlation suggests that the entire \ion{He}{i} triplet emerges from a single region, likely the stellar chromosphere.
The autocorrelation matrix of Pa$\beta$ (middle panel of Fig.~\ref{fig:autocorrelation}) shows a correlation / anticorrelation chessboard pattern above the noise level suggesting that the line width is slightly changing with time, getting narrower and deeper at times, then broader and shallower at some other times. In addition, both matrices show that the variability is asymmetric, being larger in the blue wing of these lines, possibly indicating the presence of a stellar wind. No particular pattern is apparent in the autocorrelation of the Br$\gamma$ line (right panel of Fig.~\ref{fig:autocorrelation}). We also show the normalized autocorrelation matrices (i.e. the $r_{ij}$ coefficients) in Fig.~\ref{fig:Pearson_autocorrelation_matrices}.

\subsection{2D Periodograms}

\begin{figure}
	\centering
    \begin{subfigure}[b]{0.45\textwidth}
         \centering
         \includegraphics[scale=0.27, trim= 0cm 0cm 2cm 1 cm, clip]{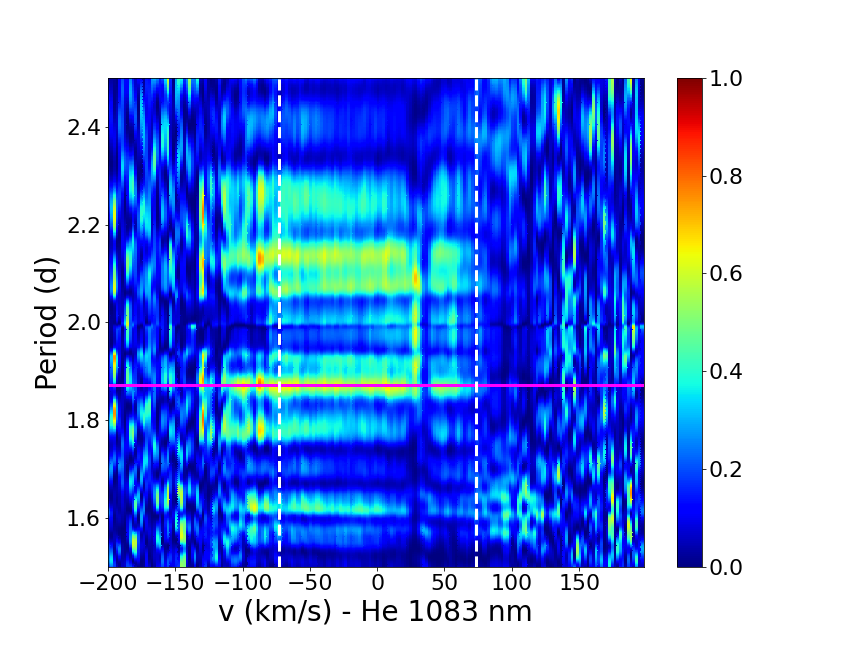}
    \end{subfigure}
    \hfill
    
	\caption{2D periodogram for the \ion{He}{i} line at 1083.3~nm obtained using the \texttt{PyAstronomy PYTHON} module \citep{pyastronomy}. A Generalized Lomb-Scargle periodogram \citep{zechmeister09} has been computed for each velocity bin and represented through a color code. The color reflects the power of the associated period in the periodogram, normalized to 1 (red indicates most powerful periods). The magenta solid line depicts the stellar rotation period while the vertical dashed lines correspond to $\pm v\sin{i}$. The periodogram highlights a period compatible with the stellar rotation period but also aliases associated with the observing window. We note that the velocity bins between 20 and 40~\kms\ were poorly corrected from telluric lines which affected the periodograms.} 
	\label{fig:he_periodogram}
\end{figure}

For each velocity bin of the median-divided spectra, we computed a Generalized Lomb-Scargle periodogram as introduced by \cite{zechmeister09} thanks to the \texttt{PyAstronomy PYTHON} module \citep{pyastronomy}. We show them as 2D maps in Fig.~\ref{fig:he_periodogram} and \ref{fig:periodogram_PaB_BrG}. Using the typical error bars derived in Sec.~\ref{sec:activity_index} for all spectral points, we found that the \ion{He}{i} and Pa$\beta$ profiles exhibit rotational modulation (with aliases associated with the observing window), with a stronger variability in the blue wing (as seen in Sec.~\ref{sec:sec4.3}). These results are consistent with those obtained from the EWVs in Sec.~\ref{sec:activity_index}, though less obvious as information is not integrated over the line profile. The Br$\gamma$ periodogram does not show any clear period, consistent with EWV$\unsim0$.

\section{Summary and discussion}
\label{sec:sec5}

Our paper reports new results derived from spectropolarimetric and photometric observations of the wTTS V410~Tau collected with the NIR spectropolarimeter SPIRou from 2019 October 31 to December 13 and the TESS space probe, from 2019 November 28 to December 23.

\subsection{Benefit of photometry}
\label{sec:5.1}
For the first time, we jointly used high-resolution spectropolarimetry and high-precision photometry in ZDI to reconstruct the brightness distribution at the surface of the star. Both data are complementary as spectropolarimetry mostly constrains the location of the spots (from the distorsions of profiles) while photometry mainly informs on their contrast relative to the quiet photosphere. 

Taking into account photometry yields a higher number of brightness features and higher contrasts in the ZDI image, especially at low latitudes (Fig.~\ref{fig2}). These features are needed so that both SPIRou and TESS data can be fitted at unit \chisqr\, with the light curve fitted down to 1.6~mmag RMS (with residual correlated noise likely attributable to small-scale rapidly-evolving surface brightness features that cannot be properly reproduced by ZDI). These latitudes are usually not well reconstructed when considering spectropolarimetry only, as ZDI is best sensitive to features located at higher latitudes, in the visible hemisphere \citep{Vogt87,brown91}. We suspect that this increase in spot coverage reflects that TESS and SPIRou do not see the same spot distributions because of the difference of spectral domains. To further improve the accuracy of the brightness modeling at the surface of the star, one would need to secure photometric data in the $JHK$ bands that would provide an ideal match to the SPIRou spectropolarimetric data. This would ensure in particular that all brightness features present at the surface of the star (including those in the polar regions) affect spectroscopic and photometric data in the same way (which is not the case for, e.g., the cool polar spot detected in the optical but not in the NIR).

Ground-based photometry allowed us to perform a similar analysis as \cite{yu19} (Fig.~B7 of their paper) but with $V-R_{\rm c}$ and $V-I_{\rm c}$ colour indexes. We fitted these indexes as a function of the magnitude in the $V$ band using a simplistic two-temperature model based on colour indexes from \cite{bessel98}. Our model features a fixed temperature of 4500~K for the photosphere, a surface gravity $\log{g}=4.0$, and a fixed temperature for spots with varying filling factor. We found an optimal spot temperature of 3750~K with a typical surface spot coverage of about 70\% (Fig.~\ref{fig:two_temperatures_model}), consistent with previous photometric measurements \citep{yu19}. This rather high spottedness level (consistent with that found for the similar wTTS LkCa~4, \citealt{gullysantiago17}) suggests in particular that a large fraction of the stellar surface is more or less evenly covered with small features that are not accessible to (and thus not reconstructed by) ZDI.

\subsection{Infrared vs optical brightness reconstruction}

V410~Tau is a wTTS that has been extensively studied in the past, mainly in the optical. Our study is innovative as we used NIR observations to constrain the brightness and the magnetic field of the star.

As expected, NIR leads to a less contrasted surface brightness map. We find a reasonable statistical agreement for low- and mid-latitude spots (but less so in the polar regions) between optical and NIR maps, even though secured at different epochs. However, the absence of a polar spot in the map reconstructed from NIR data is surprising since such a feature consistently showed up in images derived from optical data up to now \citep{joncour94,hatzes95,skelly10,rice11,caroll12,yu19}. Our reconstructed map obtained with a mask containing only molecular lines, more sensitive to cooler regions, further confirms that no polar spot is detected at NIR wavelengths. Although we cannot entirely dismiss it, the option that the polar spot disappeared at the time of our SPIRou observations seems unlikely given the persistent presence of this feature in all previously published studies.  The fact that the TESS light curve yields an average rotation period that is consistent with the trend derived from previous photometric data by \cite{yu19} suggests that the spot configuration at the surface of V410~Tau did not drastically evolve since 2016 and in particular that the cool spot reconstructed near the pole from optical data was likely still present in 2019. 

If the dark polar feature systematically seen at optical wavelengths is indeed not visible in the NIR, it suggests that continuum opacity above polar regions of V410 Tau is much larger in the optical domain than in the NIR for some reason. A speculative option, to be investigated further, may be that dust grains, such as those present in the upper solar atmosphere though in larger concentrations, tend to cluster in polar regions of the upper atmosphere of V410~Tau, making them appear much darker at optical than at NIR wavelengths.

\subsection{Magnetic field}

Applying ZDI to our Stokes~$I$ and $V$ LSD profiles simultaneously allowed us to reconstruct the large scale magnetic topology of the star. Our results are consistent with previous studies \citep{skelly10,yu19}. We find that the large-scale magnetic field has an average surface strength of about 410~G and that the radial field can be more intense locally, reaching up to 1.1~kG. Although V410~Tau is still fully convective, the magnetic field presents a strong toroidal component of unclear origin, as for the other fully-convective wTTS LkCa4 \citep{donati14}. More observations of fully-convective wTTSs are thus clearly needed to further constrain the origin of this strong toroidal field.

We also found that the poloidal component contributes to nearly 60\% of the overall magnetic energy, compatible with the recent measurements derived from NARVAL (at the Telescope Bernard Lyot) optical data in 2016 \citep{yu19}. In addition, the polar strength of the dipole component (of the poloidal field) is close to 400~G, which again supports the reported increase in the intensity of the dipole from 2008 \citep{yu19}. These properties are compatible with those obtained by \cite{yu19} from their 2016 data set, and more generally with the long-term evolution they pointed out. These results suggest that, if a magnetic cycle exists, it is likely longer than 11~years. More observations of V410~Tau would be needed to confirm whether the observed tendency reflects part of a magnetic cycle as suggested by other studies \citep{stelzer03,hambalek19} or rather intrinsic variability of a stochastic nature. 

The longitudinal field as derived from SPIRou data shows similar fluctuations than that from optical studies (of period $\unsim P_{\rm rot}$), but with error bars that are about 1.7 times smaller (typically 30~G) in half the exposure time, clearly demonstrating the benefits of studying magnetic fields of young stars in the NIR thanks to the enhanced Zeeman effect.

We constrained the surface differential rotation of V410~Tau with ZDI from our Stokes~$I$ and $V$ LSD profiles separately, both results being compatible within 1.5$\sigma$. Our estimates are also consistent with those provided by \cite{yu19} within $\unsim3 \sigma$, although our Stokes~$I$ LSD profiles yielded slightly lower value. We note that the error bars obtained from Stokes~$I$ LSD profiles are larger in the NIR than in the optical, which is likely due to the lower number of observations but also to the lower contrast of the brightness features. For Stokes V data, the differential rotation parameters we derived are similar to the optical measurements of Yu et al (2019), with error bars of the same magnitude despite the sparser data thanks to the enhanced Zeeman effect in the NIR.

We note that our estimates of differential rotation are larger than those derived by \citealt{siwak11} from photometric data collected with the MOST space-telescope in 2009. This photometric measurement of differential rotation is also inconsistent with the estimates of \cite{yu19}, despite having been collected at a close-by epoch. We thus suspect that this difference is related to the two-spot model used by \cite{siwak11} known to be inappropriate for stars like V410~Tau given the complex spot distributions reconstructed with ZDI (featuring both bright and dark spots).

\subsection{Chromospheric activity}

The \ion{He}{i} triplet at 1083~nm, the Pa$\beta$ and Br$\gamma$ lines are used as proxies to study the chromospheric activity of V410~Tau. A flare was detected at cycle 21.752 which also affected the two subsequent observations. Our analyses reveal that both the \ion{He}{i} and the Pa$\beta$ lines are rotationally modulated, while no significant variations are observed in Br$\gamma$.

\begin{figure}
    \centering
    
     \begin{subfigure}[b]{0.23\textwidth}
         \centering
         \includegraphics[scale=0.05]{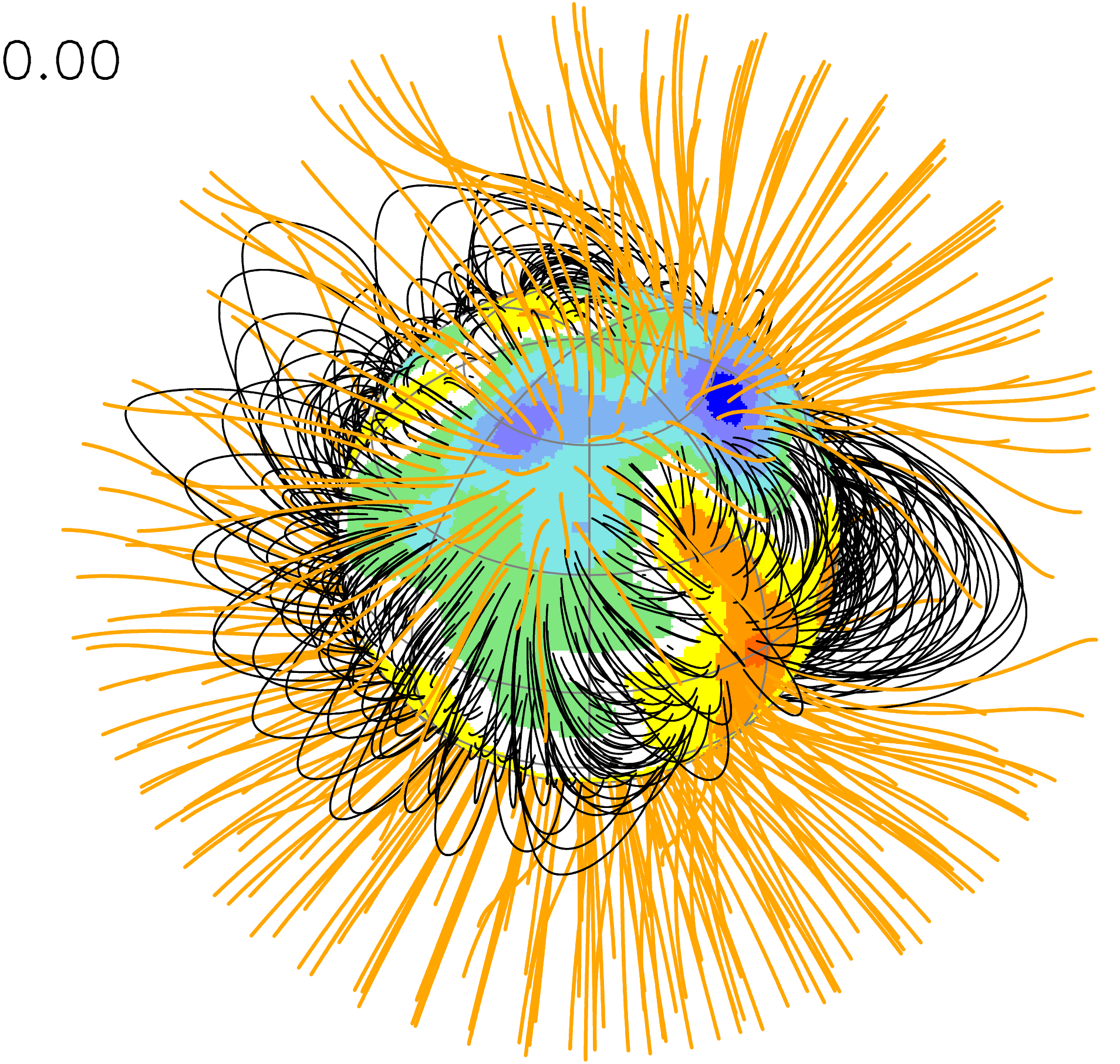}
         \label{fig:phase00}
    \end{subfigure} 
    \begin{subfigure}[b]{0.23\textwidth}
         \centering
         \includegraphics[scale=0.05]{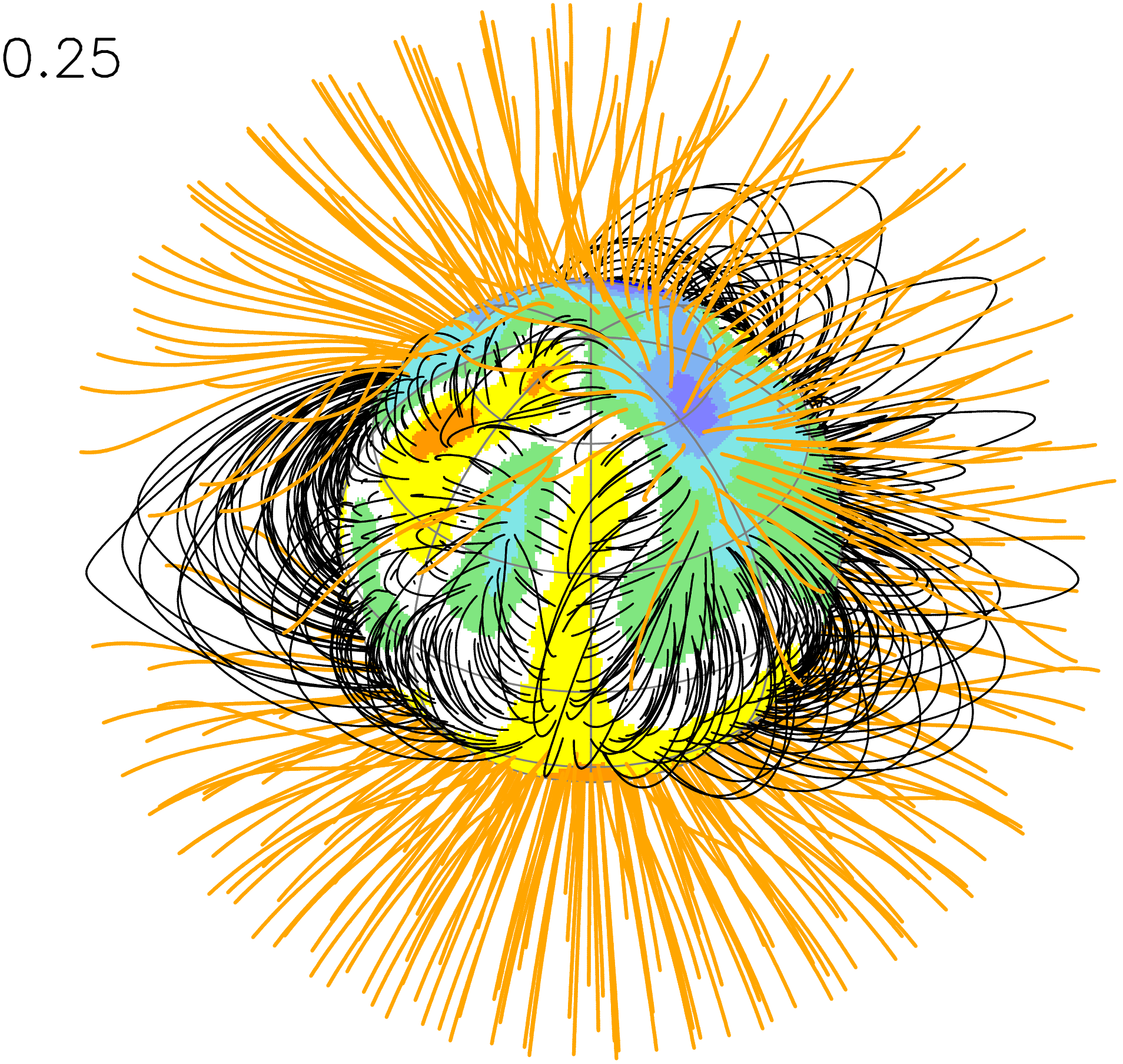}
         \label{fig:phase20}
    \end{subfigure}
    \begin{subfigure}[b]{0.23\textwidth}
         \centering
         \includegraphics[scale=0.05]{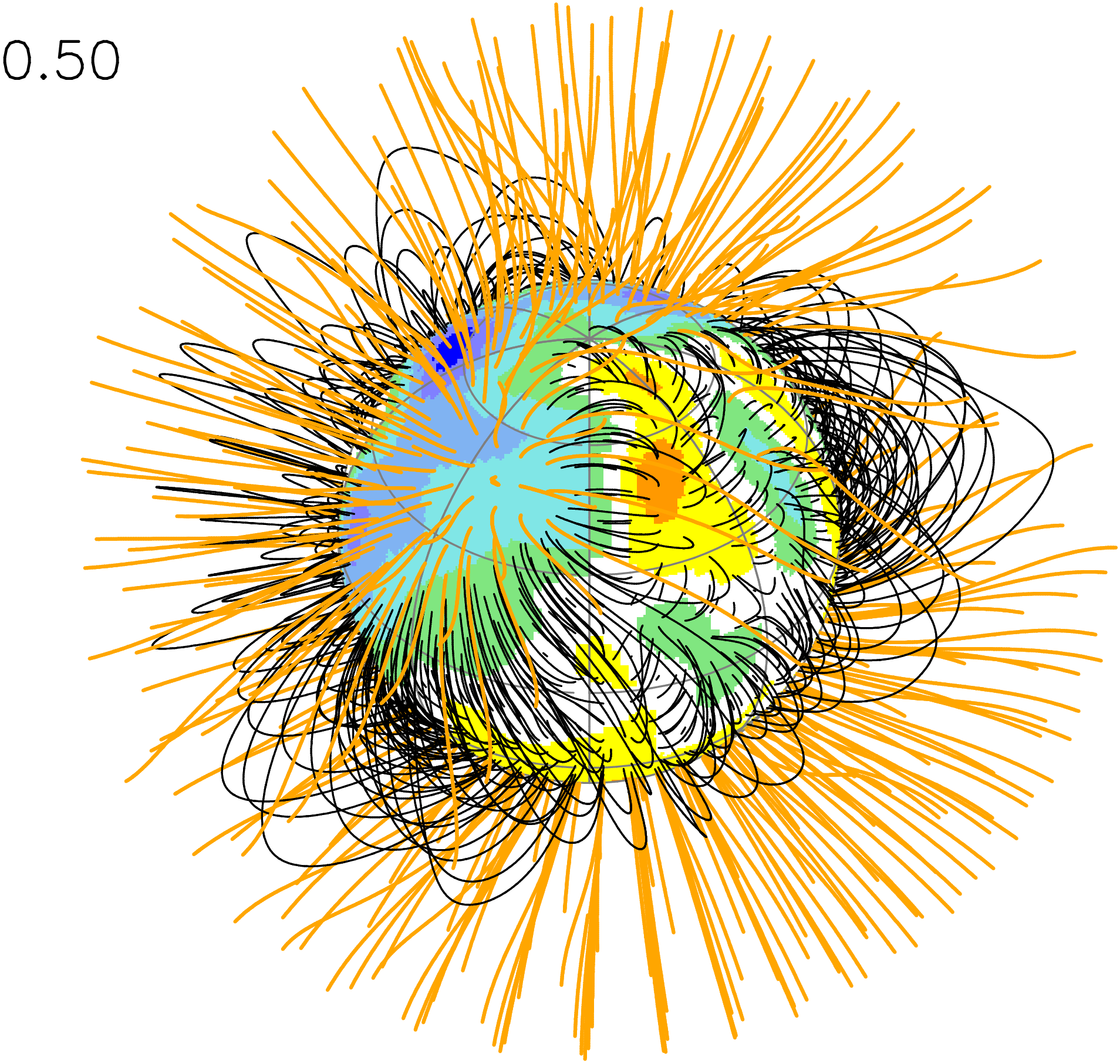}
         \label{fig:phase50}
    \end{subfigure}
    \begin{subfigure}[b]{0.23\textwidth}
         \centering
         \includegraphics[scale=0.05]{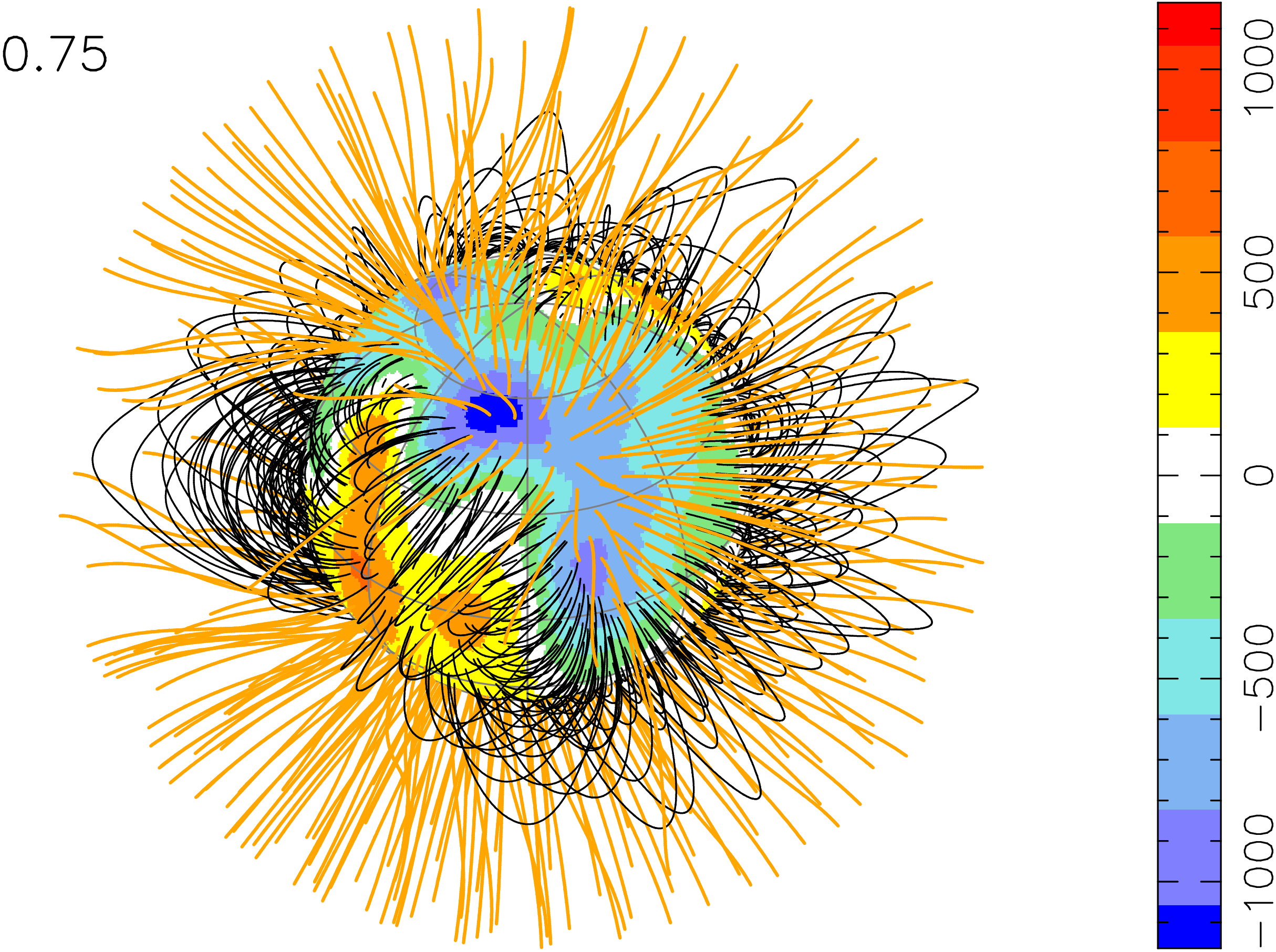}
         \label{fig:phase70}
    \end{subfigure}
 
    \caption{Potential field extrapolations of the surface radial magnetic field obtained with ZDI, as seen by an Earth-based observer. Open field lines are shown in orange while closed field lines are drawn in black. Colours at the stellar surface represent the local value of the radial magnetic field (in G). Following \citet{yu19}, we assumed that the source surface is located at 2.1~R$_\star$, corresponding to the corotation radius, beyond which field lines open under the impact of centrifugal force. The star is shown at 4 evenly-spaced phases of the rotation cycle (indicated in the top left corner of each plot). }
    \label{fig:extrapolated_field}
\end{figure}

To obtain a rough description of the large-scale stellar magnetosphere, we extrapolated our magnetic image at the surface of V410 Tau into 3D maps, assuming that the magnetic field is potential (following the method described by \citealt{jardine99}) and that the source surface at which field lines open is located at 2.1~R$_\star$, following \cite{yu19} (Fig.~\ref{fig:extrapolated_field}). We see that enhanced absorption in chromospheric lines, occuring in phase range 0.4-0.6, takes place slightly before the magnetic pole crosses the line of sight (at phase 0.7, see Fig.~\ref{fig:extrapolated_field}), i.e., when one may have expected it to occur by analogy with solar coronal holes (darker in regions of open field lines). This phase lag may relate to the potential field assumption being no more than a rough approximation in our case. The reconstructed large-scale magnetic field indeed features a strong toroidal component (with intense azimuthal fields located close to the open field line region at phase 0.7, see Fig.~\ref{fig:maps_Iv}) that may suggest that the large-scale surface field is significantly stressed at these phases. Another option is that this enhanced absorption episode is due to the presence of massive prominences trapped in closed coronal loops (such as those reported in \citealt{yu19}) and crossing the stellar disc at phases 0.4-0.6.  

Obviously, the way the \ion{He}{i} triplet and the Pa$\beta$ line behave in wTTSs, and in particular how the \ion{He}{i} and Pa$\beta$ fluxes respond to the topology of the large-scale field remains to be investigated in more details. This will be the subject of forthcoming papers.

\subsection{Filtering activity jitter from RV curves}
\label{sec:sec5.5}

RVs of V410 Tau derived from Stokes~$I$ LSD profiles exhibit a full amplitude of about 4.5~\kms\ and a dispersion of 1.40~\kms\ RMS. These values are smaller than those generally observed in the optical at roughly the same SNRs, with amplitudes ranging from 4 to 8.5~\kms\ and a typical dispersion of 1.8~\kms\ RMS, respectively \citep{yu19}. This confirms the gain in using NIR observations to reduce the activity jitter in RV measurements. We find that the amplitude of RV jitter is reduced by up to a factor of 2, consistent with results of previous optical and NIR RV studies of TTSs \citep{prato08,mahmud11,crockett12}.

For each of our ZDI reconstructed brightness maps (Fig.~\ref{fig2}) we computed the RV curve that results from the brightness features at the surface of V410~Tau. The ZDI image taking into account SPIRou data only, corresponding to a static brightness distribution, yields filtered RVs with a dispersion of 0.13~\kms\ RMS, i.e. about 25\% lower than in the optical for this star (typically 0.17~\kms; \citealt{yu19}), which suggests that the evolution of spots is not significant over our observations. We note that adding TESS data to the SPIRou data in the ZDI modeling does not improve, and actually even degrades, the accuracy of the filtering process (even after explicitly taking into account the difference in brightness contrasts at SPIRou and TESS wavelengths in the imaging process). This result demonstrates that the brightness distributions as seen by SPIRou and TESS are genuinely different and cannot be simply scaled up from one another, e.g., using Planck's law, with some features showing up in one spectral domain but not in the other (like the prominent polar spot detected ine the optical but not seen at NIR wavelengths). Filtering the activity jitter would thus likely be more efficient with ZDI applied to data sets combining SPIRou data with high-precision NIR photometry.
We also modeled the RV activity jitter using GPR yielding a dispersion of filtered RVs about twice smaller than with ZDI models thanks to the higher flexibility of GPR to model intrinsic variability in the periodic modulation of the RV curve, that results from the evolution of the spot configuration at the surface of the star. 

The periodograms of RVs (Fig.~\ref{fig:Filtered_RV_periodogram}) do not show any periodic signature beyond that from V410~Tau~A, which further confirms that our spectropolarimetric data mainly probe the primary star, and not (or no more than very marginally) its 2 companions. In addition, our filtered RVs show no evidence for a RV signal from a potential giant planet on a close-in orbit (Fig.~\ref{fig:Filtered_RV_periodogram}), consistent with previous observations that did not suggest the presence of a hJ \citep{yu19}. To derive an upper limit on the mass of a potential planet from our data, we proceeded as in \cite{yu19} and applied GPR on simulated datasets (with the same temporal sampling as that achieved for our 2019 observations) featuring both a RV activity jitter (computed from the results of Sec.~\ref{sec:sec4.1}) and a RV signal from a planet on a circular orbit with a white noise identical to that of our measurement (of 181~\ms\ RMS). For each simulation, we compared models including both the planet and the activity jitter, with those including only the activity jitter, to assess the significance level at which a close-in giant planet of given mass could be detected from our data. From the difference of logarithmic marginal likelihood between both models (detection threshold set at $\Delta \mathcal{L}=10$), we found that, for a planet-star separation lower than 0.09~au, only planets with a mass larger than $\unsim5$~M$_{\rm jup}$ can be reliably detected (at a >3$\sigma$ level), consistent with the (more stringent) upper limit (of $\unsim1$~M$_{\rm jup}$) derived by \cite{yu19} for the same planet-star separation. 

Detecting close-in massive planets typically requires carrying out monitorings over several months during which the surface of the star can evolve significantly. This intrinsic variability cannot be modeled with the current version of ZDI that assumes a static distribution of features at the surface of the star (except for differential rotation), often forcing one to split data sets into smaller subsets that can be modelled independently from one another (e.g. \citealt{donati17,yu17,yu19}). In order to get a more global and consistent description of the stellar surface activity over several months, one needs to be able to model at the same time both the distribution of surface (brightness and magnetic) features and its evolution with time using all data at once. In this aim, we started to modify the original ZDI code to couple it with GPR, in order to simultaneously benefit from the physical modeling provided by ZDI (to detect and characterize stellar surface features), and from the flexibility provided by GPR (to describe the temporal evolution of these features). This new version of ZDI is currently under development and will be the object of forthcoming publications.

Our study illustrates the benefits of NIR (vs optical) observations with instruments like SPIRou, to investigate the magnetic topologies of young stars and look for the potential presence of hJ on close-in orbits through RV measurements. New monitorings of V410~Tau will provide strong constraints on the existence of a magnetic cycle (and the underlying dynamo processes), will bring further clues on the enigmatic strong toroidal field that the star is able to trigger despite being fully convective, and will allow us identifying the main differences between images reconstructed from optical and NIR data, especially in the polar regions. More generally, SPIRou observations of PMS stars, including those carried out within the SLS, will offer the opportunity to investigate in more details the impact of magnetic fields on star / planet formation, and in particular to accurately characterize young planetary systems hosting transiting planets such as AU Mic and V1298 Tau \citep{david19a,david19b,plavchan20,klein21}, allowing one to refine the mass-radius relation of planets at an early stage of evolution.

\section*{Acknowledgements}

This work includes data collected with SPIRou in the framework of the SPIRou Legacy Survey (SLS), an international large programme allocated on the Canada-France-Hawaii Telescope (CFHT), operated from the summit of Maunakea by the National Research Council of Canada, the Institut National des Sciences de l’Univers of the Centre National de la Recherche Scientifique of France, and the University of Hawaii. We acknowledge funding by the European Research Council (ERC) under the H2020 research \& innovation programme (grant agreements \#740651 NewWorlds, \#865624 GPRV and \#716155 SACCRED). SHPA acknowledges financial support from CNPq, CAPES and Fapemig. We thank the referee for valuable comments and suggestions that helped improving the manuscript.

\section*{Data Availability}

The data collected with the TESS space probe are publicly available from the Mikulski Archive for Space Telescopes (MAST).
The SLS data will be publicly available from CADC one year after the completion of the SLS programme (in 2022). 

\bibliographystyle{mnras}
\bibliography{paper_v410}

\appendix

\section{Stokes I LSD profiles obtained with the M3 mask}

\begin{figure}
    \centering
    \includegraphics[scale=0.15]{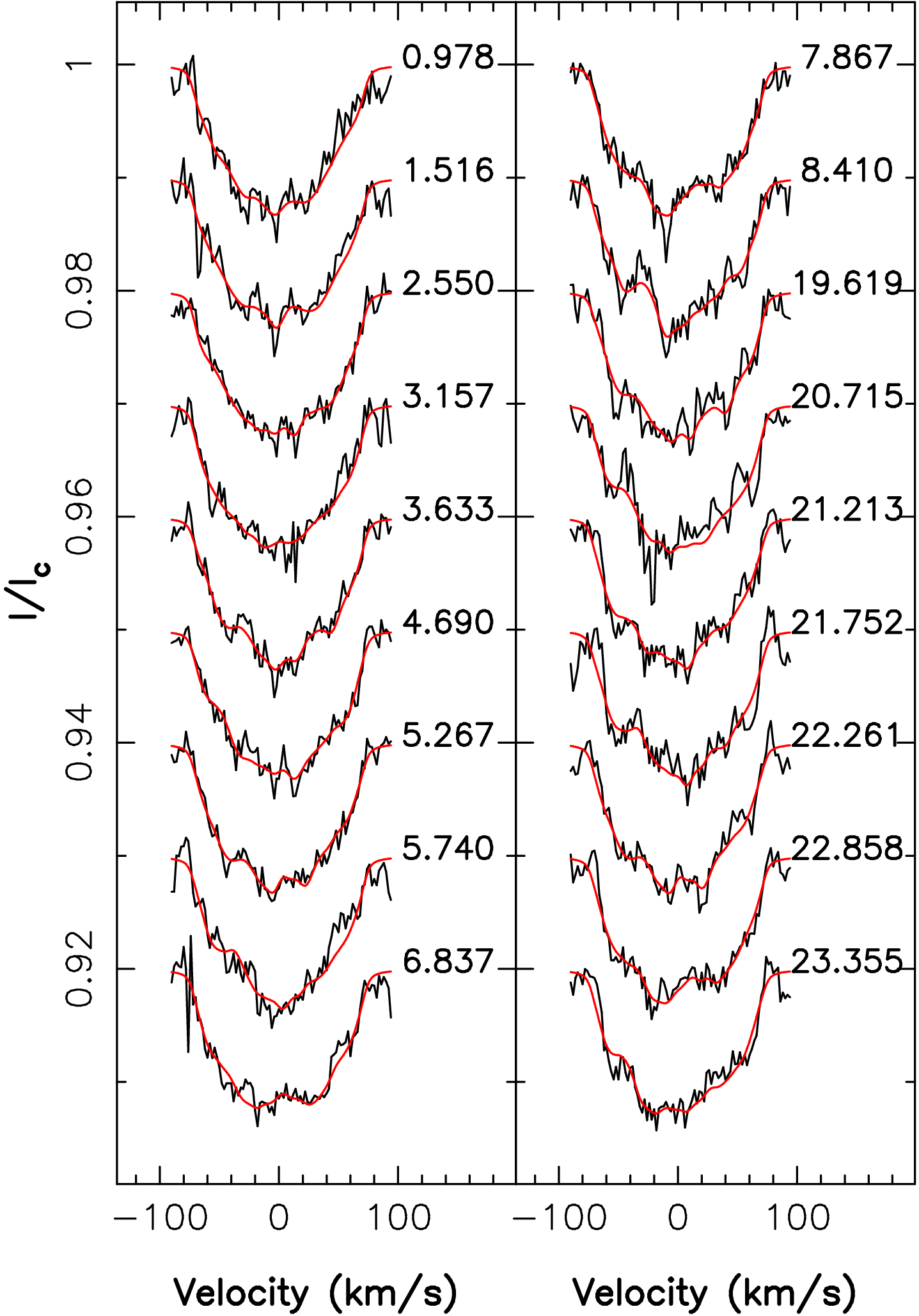}
    \caption{Stokes~$I$ LSD profiles obtained with a mask containing only molecular lines. The observed profiles are plotted in black while the ZDI model is plotted in red, with the associated rotation cycle mentioned on the right.}
    \label{fig:StokesI_M3}
\end{figure}

We show the Stokes~$I$ LSD profiles obtained with the M3 mask, containing only molecular lines, in Fig.~\ref{fig:StokesI_M3}.

\section{Activity proxies}

\subsection{Spectra}

\begin{figure}
    \centering \hspace*{-1.1cm}
    \begin{subfigure}[b]{0.49\textwidth}
         \centering
         \includegraphics[scale=0.33]{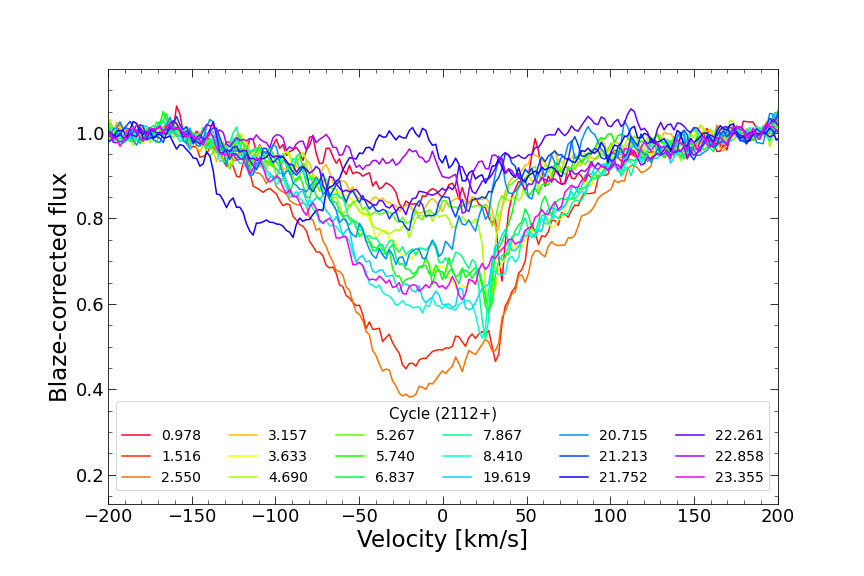}
         
    \end{subfigure}
   
    \hspace*{-1.1cm}
    \begin{subfigure}[b]{0.49\textwidth}
         \centering
         \includegraphics[scale=0.33]{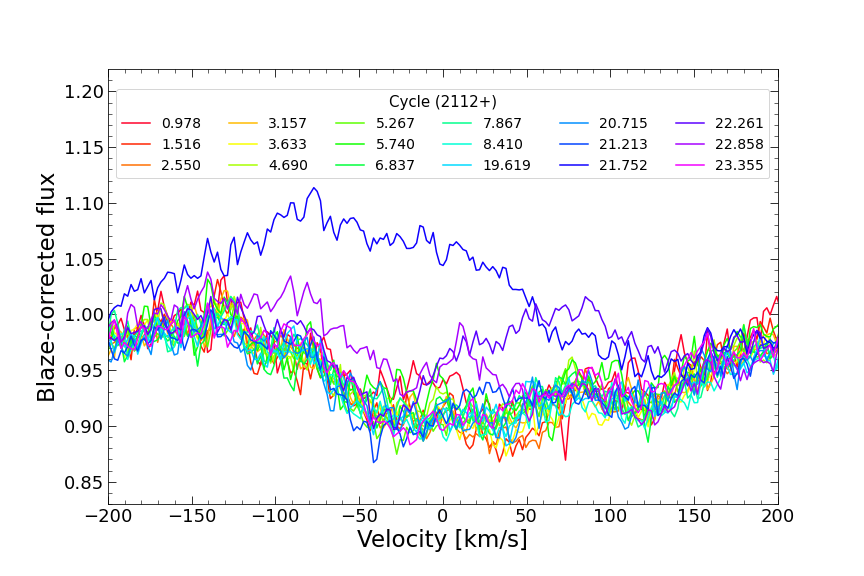}
       
    \end{subfigure}
    
    \hspace*{-1.1cm}
    \begin{subfigure}[b]{0.49\textwidth}
         \centering
         \includegraphics[scale=0.33]{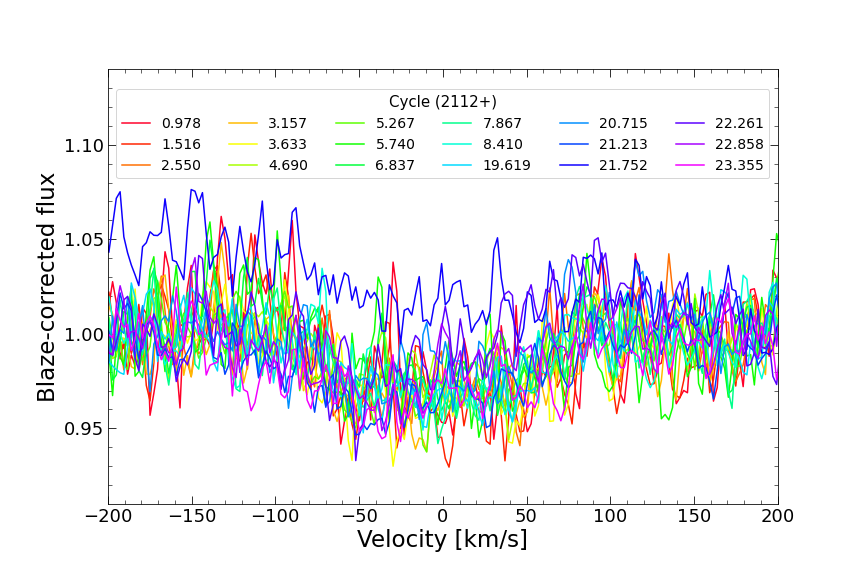}
         
    \end{subfigure}
    \caption{Observed profiles for \ion{He}{i} (top panel), Pa$\beta$ (middle panel) and Br$\gamma$ (bottom panel) lines. The shallow depression around 120~\kms\ in Pa$\beta$ is likely related to \ion{Ti}{}, \ion{Ca}{}, and \ion{Fe}{} lines that blend with the red wing of Pa$\beta$.} 
    \label{fig:spectral_lines}
\end{figure}

\begin{figure}
    \centering \hspace*{-1.cm}
    \begin{subfigure}[b]{0.49\textwidth}
         \centering
         \includegraphics[scale=0.32]{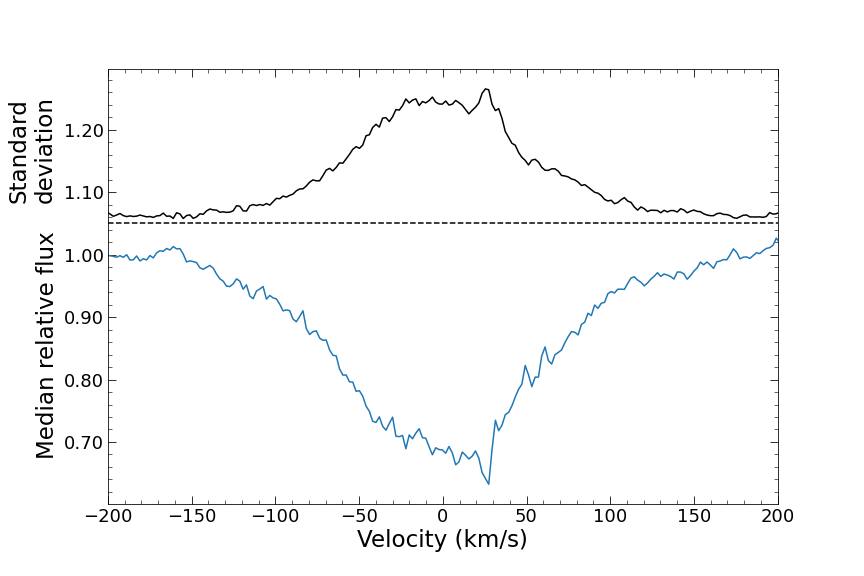}
         
    \end{subfigure}
    \hspace*{-1.cm}
    \begin{subfigure}[b]{0.49\textwidth}
         \centering
         \includegraphics[scale=0.32]{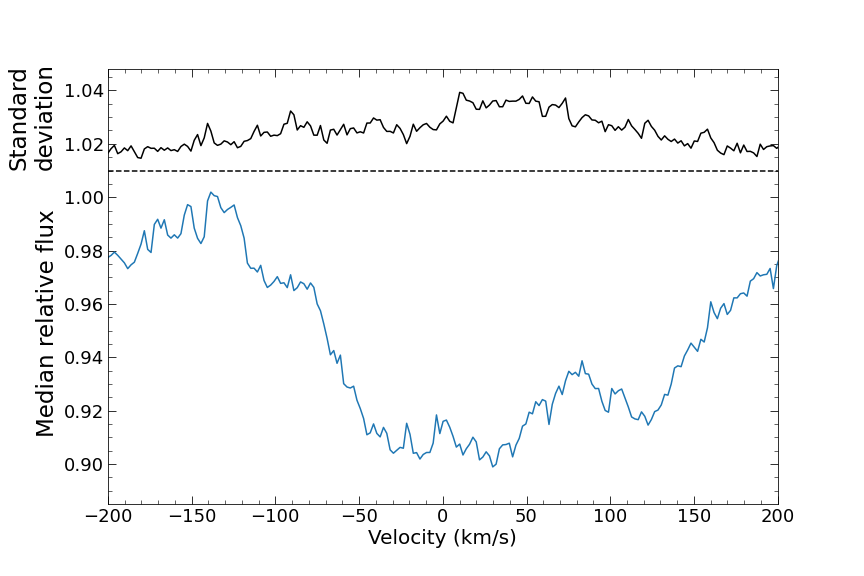}
        
    \end{subfigure}
    \hspace*{-1.cm}
    \begin{subfigure}[b]{0.49\textwidth}
         \centering
         \includegraphics[scale=0.32]{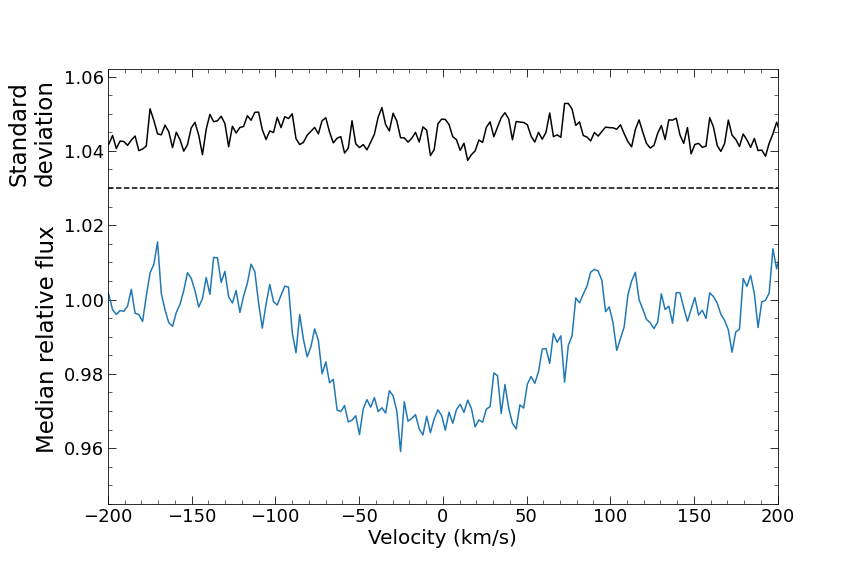}
         
    \end{subfigure}
    \caption{Median profiles (blue/bottom curves) and dispersion in the velocity bins of the median-divided spectra (black/top curves), computed after removing the observations affected by a flare, for \ion{He}{i} (top panel), Pa$\beta$ (middle panel) and Br$\gamma$ (bottom panel) lines. The black curves are shifted upwards by 1.05, 1.01 and 1.03 for the \ion{He}{i} triplet, Pa$\beta$ and Br$\gamma$ lines, respectively, for display purposes (the dashed line thereby depicting the zero variability level). Pa$\beta$ is blended with Ti, Ca and Fe lines in the red wing causing the shallow depression around 120~\kms. This feature does not vary more than the continuum and is thereby not expected to affect the measured activity indicators.}
    \label{fig:median_lines}
\end{figure}

\begin{figure}
    \centering \hspace*{-1.1cm}
    \begin{subfigure}[b]{0.49\textwidth}
         \centering
         \includegraphics[scale=0.33]{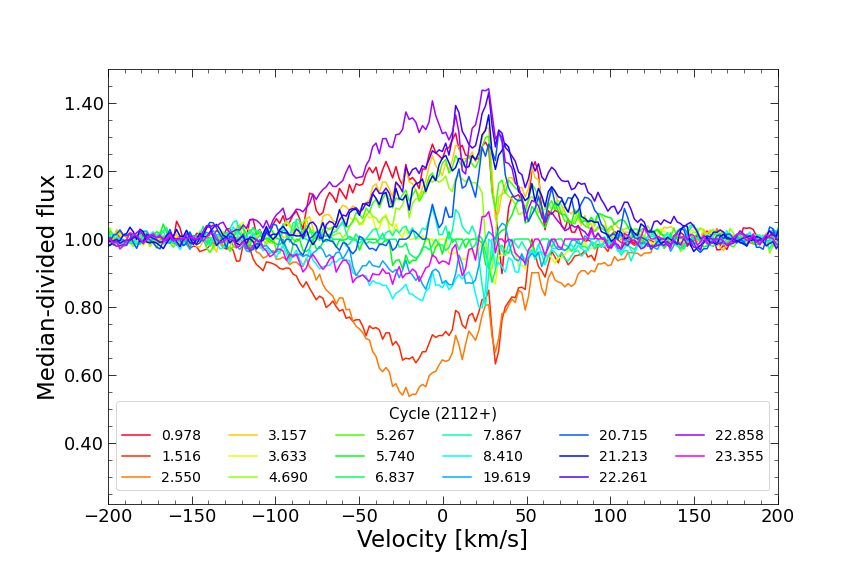}
         
    \end{subfigure}
    \hspace*{-1.1cm}
    \begin{subfigure}[b]{0.49\textwidth}
         \centering
         \includegraphics[scale=0.33]{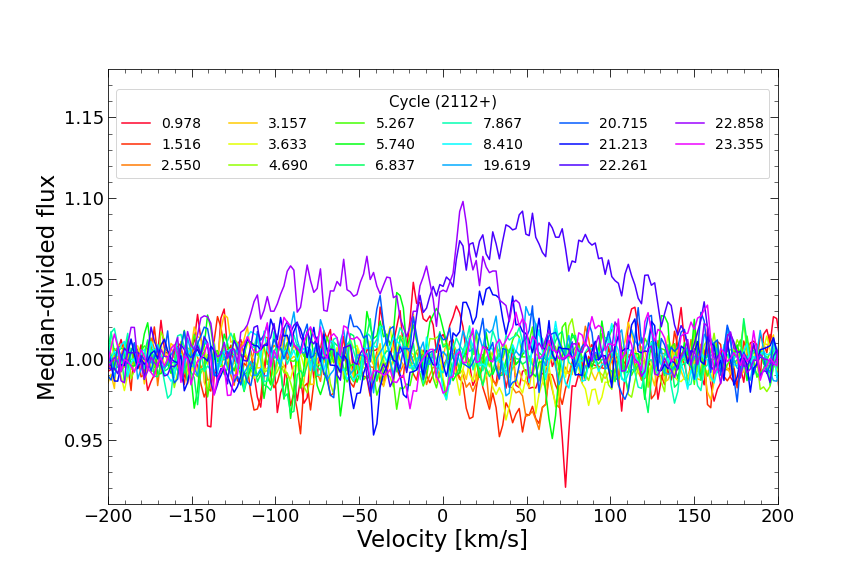}
        
    \end{subfigure}
    \hspace*{-1.1cm}
    \begin{subfigure}[b]{0.49\textwidth}
         \centering
         \includegraphics[scale=0.33]{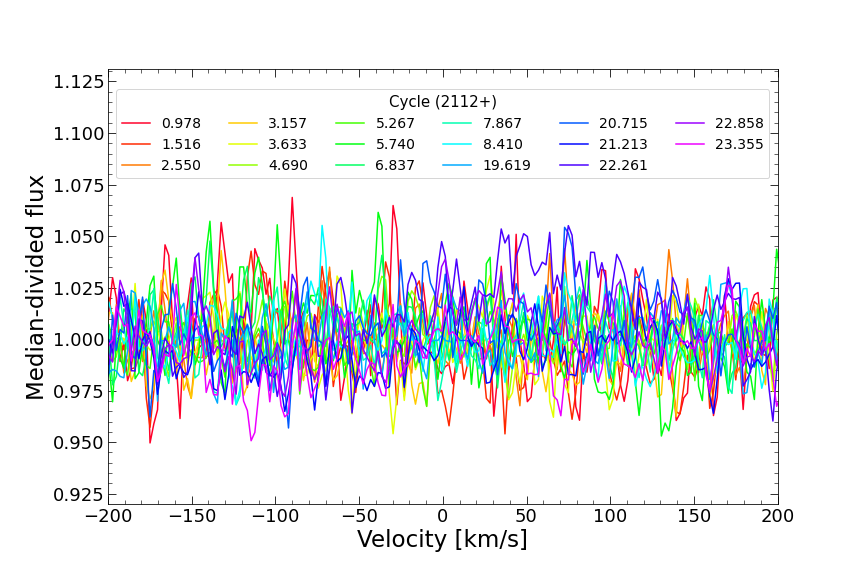}
        
    \end{subfigure}
    \caption{Median-divided profiles for \ion{He}{i} (top panel), Pa$\beta$ (middle panel) and Br$\gamma$ (bottom panel) lines after removing the profile affected by the main flare at cycle 21.752 (see Sec.~\ref{sec:activity_index}) .} 
    \label{fig:spectral_lines_residuals}
\end{figure}

The \ion{He}{i}, Pa$\beta$ and Br$\gamma$ raw profiles are shown in Fig. \ref{fig:spectral_lines}. The median profiles are shown in Fig.~\ref{fig:median_lines} while the median-divided spectra are shown in Fig.~\ref{fig:spectral_lines_residuals}. The \ion{He}{i} and Pa$\beta$ profiles exhibit enhanced emission at cycle 21.752, as well as redshifted emission at cycle 22.261 both attributed to a flare.

\begin{figure*}
    \centering
    \begin{subfigure}[b]{0.49\textwidth}
         \centering
         \includegraphics[scale=0.27]{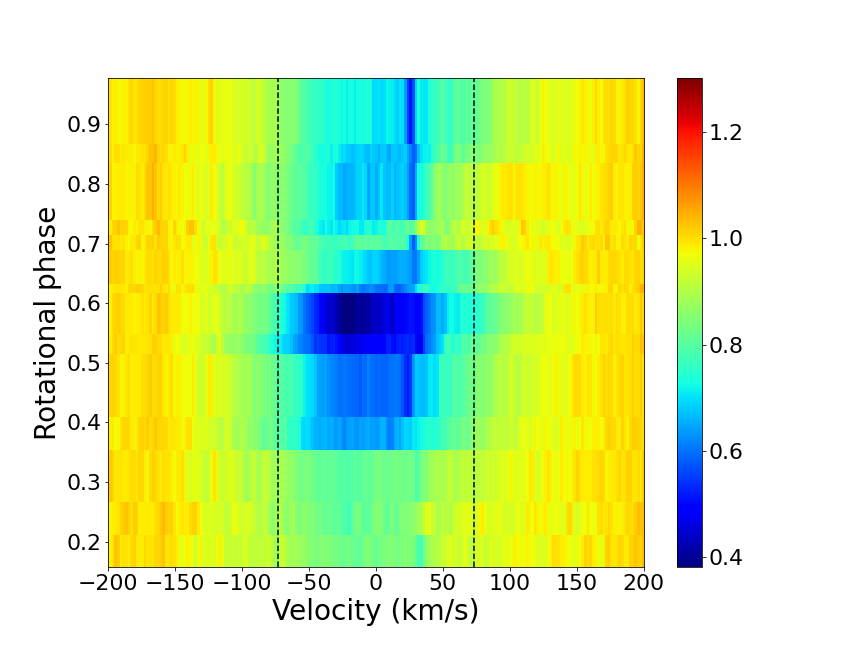}
         
    \end{subfigure}
    \hfill
    \begin{subfigure}[b]{0.49\textwidth}
         \centering
         \includegraphics[scale=0.27]{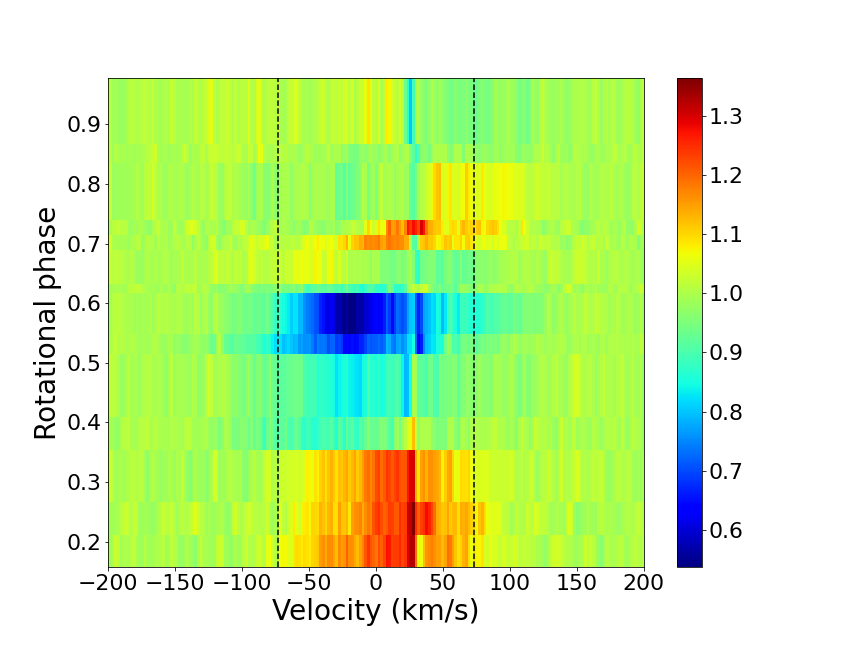}
         
    \end{subfigure}
    \caption{Dynamic spectra of \ion{He}{i} triplet (1083 nm) obtained from raw spectra (left panel) and median-divided spectra (right panel). In the left panel, the color bar refers to the intensity in the observed spectrum (blue/red meaning absorption/emission with respect to the continuum equal to 1). In the right panel, blue/red correspond to values lower/higher than the median profile. The vertical dashed lines depict $\pm v\sin{i}$. } 
    \label{fig:dynamic_spectra}
\end{figure*}

We also show the dynamic spectra of the \ion{He}{i} triplet in Fig.~\ref{fig:dynamic_spectra} (left and right panels for the raw and median-divided spectra, respectively), both exhibiting minimum emission in phase range 0.4-0.6.

\subsection{Correlation matrices}

\begin{figure*}
    \centering
    \begin{subfigure}[b]{0.3\textwidth}
         \centering
         \includegraphics[scale=0.22, trim= 0cm 0cm 2cm 1 cm, clip]{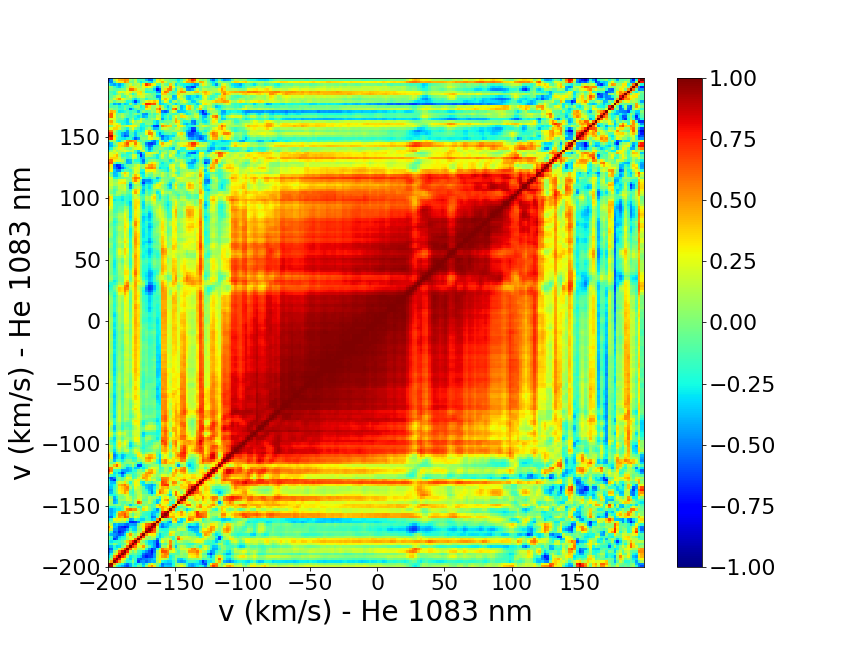}
    \end{subfigure}
    \hfill
    \begin{subfigure}[b]{0.3\textwidth}
         \centering
         \includegraphics[scale=0.22, trim= 0cm 0cm 2cm 1 cm, clip]{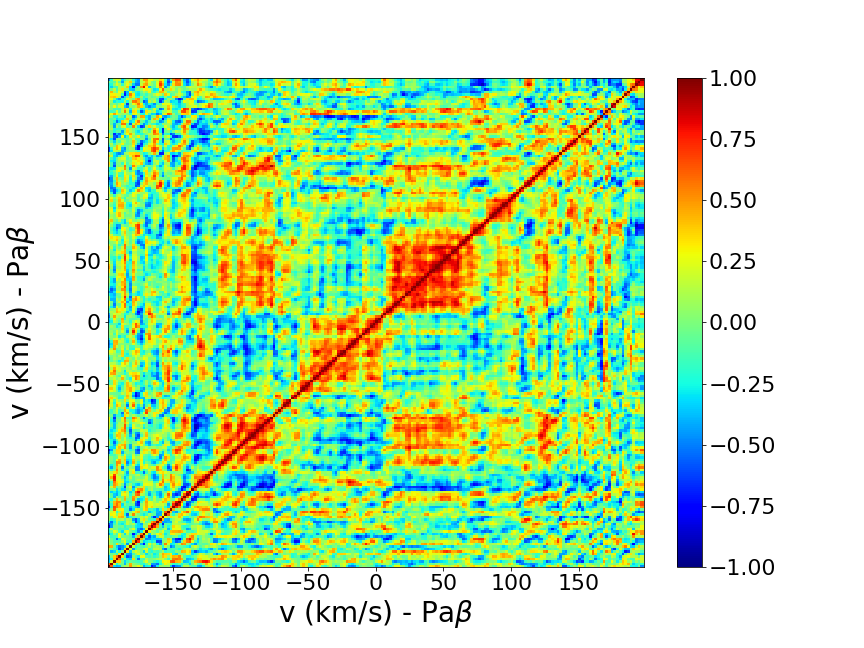}
    \end{subfigure}
    \hfill
    \begin{subfigure}[b]{0.3\textwidth}
         \centering
         \includegraphics[scale=0.22, trim= 0cm 0cm 2cm 1 cm, clip]{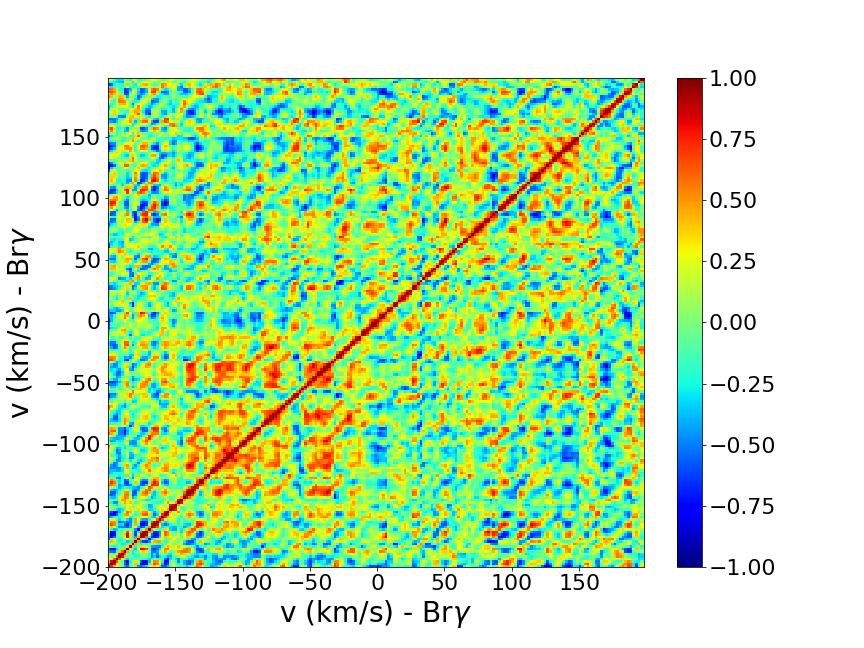}
    \end{subfigure}
    
    \caption{Normalized autocorrelation matrices of the \ion{He}{i} triplet (left), Pa$\beta$ (middle panel) and Br$\gamma$ lines (right panel). The color refers to the value of the Pearson linear coefficient going from -1 (pure anticorrelation; blue) to +1 (pure correlation; red).}
    \label{fig:Pearson_autocorrelation_matrices}
\end{figure*}

We show the normalized autocorrelation matrices representing the Pearson linear coefficient between velocity bins in Fig.~\ref{fig:Pearson_autocorrelation_matrices}.

\subsection{2D periodograms}

\begin{figure*}
    \centering
    \begin{subfigure}[b]{0.49\textwidth}
         \centering
         \includegraphics[scale=0.25, trim= 0cm 0cm 2cm 1 cm, clip]{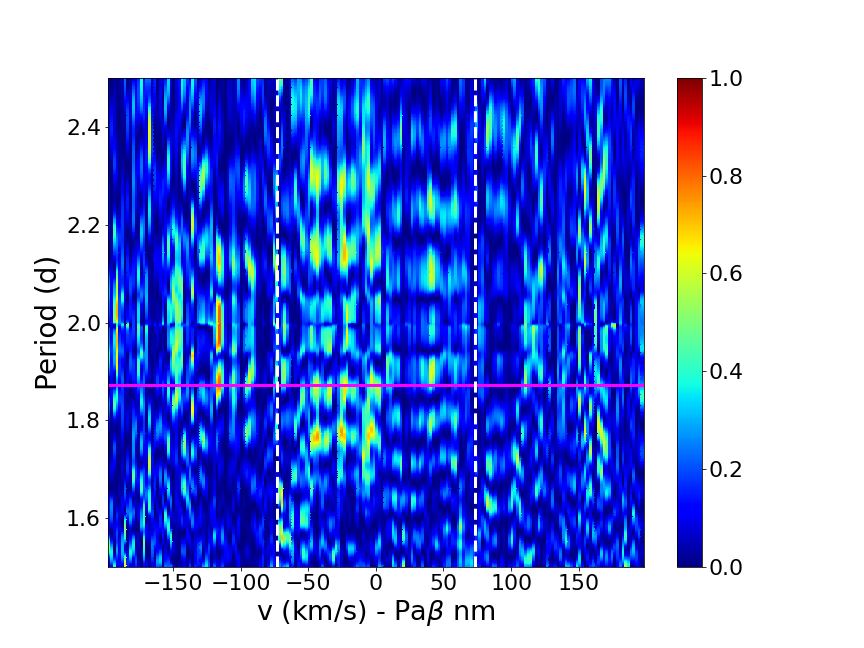}
    \end{subfigure}
    \hfill
    \begin{subfigure}[b]{0.49\textwidth}
         \centering
         \includegraphics[scale=0.25, trim= 0cm 0cm 2cm 1 cm, clip]{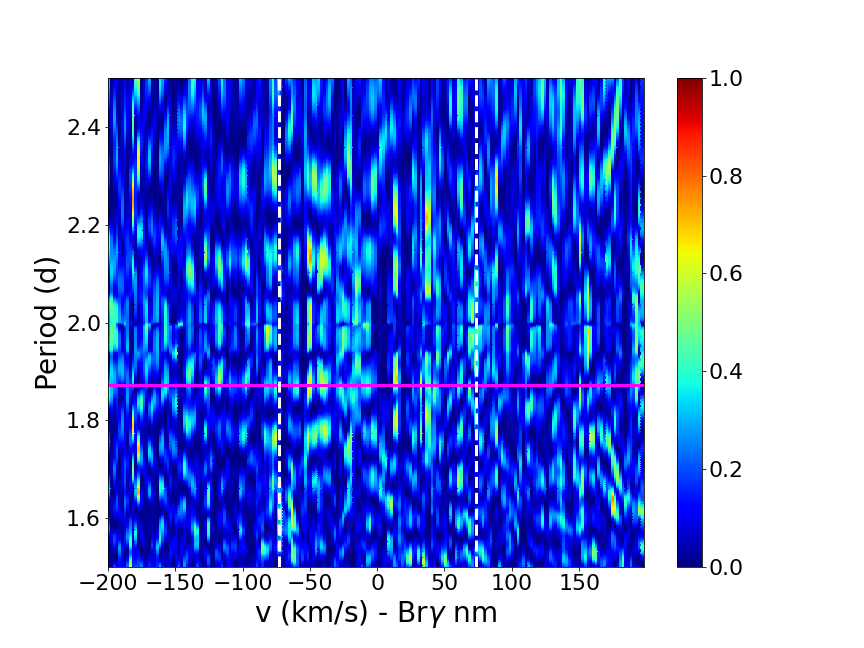}
    \end{subfigure}

    \caption{2D periodograms for the Pa$\beta$ (left panel) and Br$\gamma$ (right panel) profiles as defined in Fig. \ref{fig:he_periodogram}. The periodograms have been computed using the \texttt{PyAstronomy PYTHON} module \citep{pyastronomy}. The Pa$\beta$ periodogram highlights a period consistent with the stellar rotation period (along with aliases due to the observing window as for the \ion{He}{i} periodogram) while no clear period shows up in the Br$\gamma$ periodogram. We note that the higher peaks outside the lines likely trace stray pixels with excess residual noise. The stellar rotation period is represented by a magenta line while the vertical dashed lines depict $\pm v\sin{i}$.}
    \label{fig:periodogram_PaB_BrG}
\end{figure*}

2D periodograms for Pa$\beta$ and Br$\gamma$ profiles are shown in Fig.~\ref{fig:periodogram_PaB_BrG}. We can detect a period close to the stellar rotation period for the Pa$\beta$ periodogram but no clear period is visible in that of Br$\gamma$.

\section{Photometry}
We report information on the V410~Tau photometric data collected with the 1.25~m AZT-11 telescope at the CrAO in Table~\ref{tab:log_crao}.

\begin{figure}
    \centering
    \begin{subfigure}[b]{0.49\textwidth}
         \centering
         \includegraphics[scale=0.25]{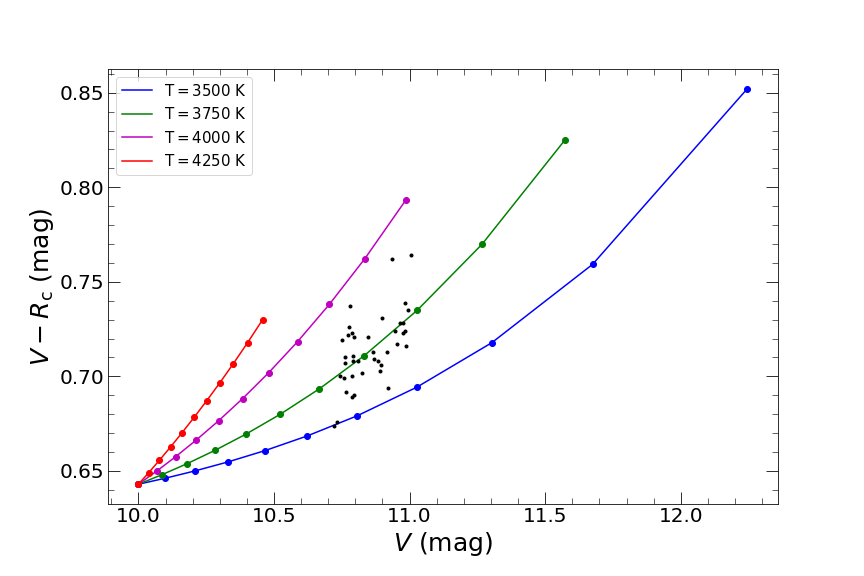}
    \end{subfigure}
    \hfill
    \begin{subfigure}[b]{0.49\textwidth}
         \centering
         \includegraphics[scale=0.25]{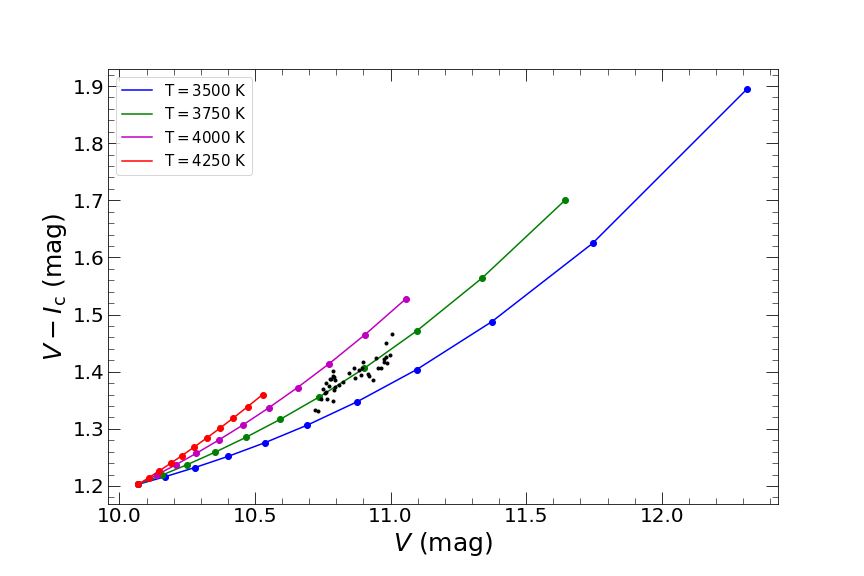}
    \end{subfigure}

    \caption{Fit of the $V-R_{\rm c}$ (left panel) and $V-I_{\rm c}$ (right panel) colour indexes as a function of the $V$ magnitude for V410~Tau in the 2019 observing season, with a two-temperature model featuring a photospheric temperature of 4500~K and synthetic colour indexes from \citet{bessel98} for $\log{g}=4.0$. Each colored line corresponds to a specific temperature for the spots. Each filled circle represents a different spot coverage, with steps of 10\% (the dots at $V=10$~mag, $V-R_{\rm c} = 0.64$~mag and at $V=10.07$~mag, $V-I_{\rm c} = 1.20$~mag corresponding to a 0\% spot coverage). Our data are shown as black dots yielding a typical spot coverage between 60 and 75\%. }
    \label{fig:two_temperatures_model}
\end{figure}

Using the synthetic colour indexes provided by \cite{bessel98}, we computed the theoretical magnitudes in the $V$, $R_{\rm c}$ and $I_{\rm c}$ bands for V410~Tau. We used these values to adjust our ground-based $V-R_{\rm c}$ and $V-I_{\rm c}$ measurements with a two-temperature model, featuring a photospheric temperature of 4500~K, a surface gravity $\log{g}=4.0$ and a fixed spot temperature. The resulting fit is shown in Fig.~\ref{fig:two_temperatures_model}. A similar analysis (with similar results) was presented in \cite{yu19}, using $B-V$ and $V$ photometric data.

\begin{table}

\caption{CrAO photometric observations of V410~Tau between 2019 September and December. The $1^{\rm st}$ and $2^{\rm nd}$ columns give the date and the Heliocentric Julian Date. In column 3, we report the measured visible magnitude ($V$). Columns 4 and 5 list the colour indexes $V-R_{\rm J}$, $V-I_{\rm J}$ in the Johnson system while columns 6 and 7 contain the colour indexes $V-R_{\rm c}$ and $V-I_{\rm c}$ in the Cousins system.}
\label{tab:log_crao}
\centering 

\begin{tabular}{lcccccc}
\\
\hline \hline
\multicolumn{1}{c}{Date} & HJD  & $V$ & $V-R_{\rm J}$ & $V-I_{\rm J}$ & $V-R_{\rm c}$ & $V-I_{\rm c}$ \\ 
\multicolumn{1}{c}{2019} &  2458700+ & (mag) & (mag) & (mag) & (mag) & (mag)  \\ \hline

Sep 02 & 29.5017 & 10.983 & 1.077 & 1.847 & 0.739 & 1.451 \\ 
Sep 03 & 30.5006 & 10.789 & 1.021 & 1.773 & 0.700 & 1.393 \\ 
Sep 05 & 32.4952 & 10.789 & 1.054 & 1.783 & 0.723 & 1.401 \\ 
Sep 07 & 34.4800 & 10.775 & 1.058 & 1.765 & 0.726 & 1.387 \\ 
Sep 08 & 35.4371 & 11.004 & 1.110 & 1.866 & 0.764 & 1.466 \\ 
Sep 10 & 37.4726 & 10.868 & 1.034 & 1.767 & 0.709 & 1.389 \\ 
Sep 12 & 39.4421 & 10.866 & 1.040 & 1.791 & 0.713 & 1.407 \\ 
Sep 13 & 40.4719 & 10.790 & 1.037 & 1.769 & 0.711 & 1.390 \\ 
Sep 25 & 52.4746 & 10.892 & 1.026 & 1.773 & 0.703 & 1.394 \\ 
Sep 28 & 55.4752 & 10.810 & 1.033 & 1.753 & 0.708 & 1.377 \\ 
Oct 02 & 59.4522 & 10.945 & 1.055 & 1.813 & 0.724 & 1.424 \\ 
Oct 03 & 60.4524 & 10.760 & 1.035 & 1.756 & 0.710 & 1.380 \\ 
Oct 06 & 63.4381 & 10.986 & 1.044 & 1.802 & 0.716 & 1.416 \\ 
Oct 08 & 65.4516 & 10.919 & 1.013 & 1.773 & 0.694 & 1.393 \\ 
Oct 10 & 67.4447 & 10.895 & 1.031 & 1.790 & 0.706 & 1.406 \\ 
Oct 12 & 69.5512 & 10.847 & 1.051 & 1.777 & 0.721 & 1.397 \\ 
Oct 13 & 70.3985 & 10.794 & 1.007 & 1.749 & 0.690 & 1.374 \\ 
Oct 20 & 77.4702 & 10.795 & 1.051 & 1.762 & 0.721 & 1.385 \\ 
Oct 22 & 79.4791 & 10.720 & 0.985 & 1.696 & 0.674 & 1.333 \\ 
Oct 24 & 81.4766 & 10.757 & 1.020 & 1.733 & 0.699 & 1.362 \\ 
Oct 26 & 83.4668 & 10.766 & 1.010 & 1.721 & 0.692 & 1.353 \\ 
Oct 27 & 84.4269 & 10.883 & 1.032 & 1.785 & 0.708 & 1.403 \\ 
Nov 01 & 89.4242 & 10.976 & 1.060 & 1.810 & 0.728 & 1.422 \\ 
Nov 02 & 90.4358 & 10.791 & 1.033 & 1.741 & 0.708 & 1.368 \\ 
Nov 03 & 91.4319 & 10.963 & 1.061 & 1.791 & 0.728 & 1.407 \\ 
Nov 05 & 93.6208 & 10.935 & 1.108 & 1.763 & 0.762 & 1.385 \\ 
Nov 09 & 97.4788 & 10.897 & 1.064 & 1.804 & 0.731 & 1.417 \\ 
Nov 10 & 98.3193 & 10.762 & 1.031 & 1.736 & 0.707 & 1.365 \\ 
Nov 16 & 104.4910 & 10.984 & 1.056 & 1.814 & 0.724 & 1.425 \\ 
Nov 17 & 105.3771 & 10.772 & 1.053 & 1.750 & 0.722 & 1.375 \\ 
Nov 18 & 106.3933 & 10.996 & 1.071 & 1.820 & 0.735 & 1.430 \\ 
Nov 19 & 107.3326 & 10.742 & 1.021 & 1.721 & 0.700 & 1.353 \\ 
Dec 02 & 120.2204 & 10.789 & 1.007 & 1.716 & 0.689 & 1.349 \\ 
Dec 04 & 122.2358 & 10.732 & 0.989 & 1.693 & 0.676 & 1.331 \\ 
Dec 05 & 123.2644 & 10.977 & 1.054 & 1.804 & 0.723 & 1.417 \\ 
Dec 06 & 124.4775 & 10.750 & 1.048 & 1.742 & 0.719 & 1.369 \\ 
Dec 07 & 125.3528 & 10.915 & 1.040 & 1.776 & 0.713 & 1.396 \\ 
Dec 08 & 126.4381 & 10.781 & 1.073 & 1.766 & 0.737 & 1.388 \\ 
Dec 14 & 132.2148 & 10.825 & 1.025 & 1.758 & 0.702 & 1.382 \\ 
Dec 18 & 136.2472 & 10.954 & 1.046 & 1.789 & 0.717 & 1.406 \\  \hline

\end{tabular}
\end{table}

\section{Radial velocities periodograms}

\begin{figure}
    \centering
    \hspace*{-.65cm}
    \includegraphics[scale=0.22, trim= 0cm 2cm 3cm 3 cm]{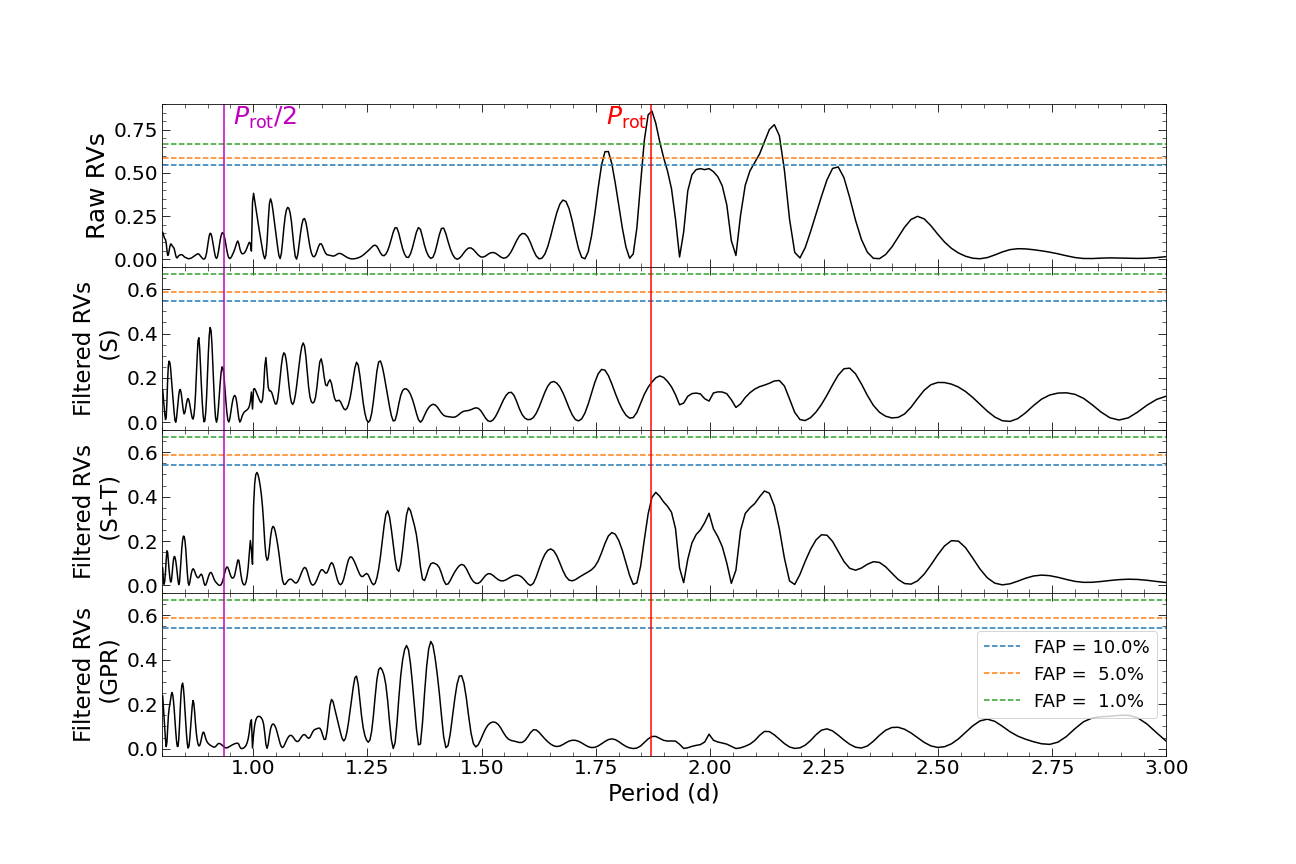}
    \caption{Periodograms of raw radial velocities (first panel) and filtered radial velocities obtained from ZDI using SPIRou (S) data only (second panel), SPIRou and TESS (S+T) data simultaneously (third panel) or GPR (fourth panel). The magenta and red vertical lines indicate $P_{\rm rot}/2$ and $P_{\rm rot}$. The horizontal dashed lines depict the FAP levels at 10\%, 5\% and 1\%. These periodograms have been computed using the \texttt{PyAstronomy PYTHON} module \citep{pyastronomy}. }
    \label{fig:Filtered_RV_periodogram}
\end{figure}

Periodograms of raw and filtered RVs are shown in Fig. \ref{fig:Filtered_RV_periodogram}. We clearly detected a period consistent with the stellar rotation period in the raw RVs but no modulation appear in all our filtered RVs suggesting that the data are not affected by stellar companions and that no close-in massive planet orbit V410 Tau A. The FAPs mentioned in Fig.~\ref{fig:Filtered_RV_periodogram} are computed assuming white noise only and may thereby be underestimated if correlated noise dominates.

\bsp	
\label{lastpage}
\end{document}